\documentclass[11pt]{article}

\usepackage{amssymb}
\usepackage{graphicx}        

\newcommand\oldcaption{}
\let\oldcaption\caption
\renewcommand\caption[2][heading]{\oldcaption[#1]{\footnotesize#2}}

\begin{document}

\begin{flushright}
\end{flushright}
\vspace{20mm}
\begin{center}
{\LARGE  What Exactly is the Information Paradox? }\\
\vspace{18mm}
{\bf  Samir D. Mathur}\\
\vspace{8mm}
Department of Physics,\\ The Ohio State University,\\ Columbus,
OH 43210, USA\\ 
\vspace{4mm}
 mathur@mps.ohio-state.edu
\vspace{4mm}
\end{center}
\vspace{10mm}
\thispagestyle{empty}
\begin{abstract}

The black hole information paradox tells us something important about the way quantum mechanics and gravity fit together. In these lectures I try to give a pedagogical review  of the essential physics leading to the paradox, using mostly pictures. Hawking's argument is recast as a `theorem': if quantum gravity effects are confined to within a given length scale and the vacuum is assumed to be unique, then there will be information loss. We conclude with a brief summary of how quantum effects in string theory violate the first condition and make the interior of the hole a `fuzzball'.

\vspace{4mm}

{\it [Prepared for the proceedings of the 4th Aegean Summer School on Black Holes, Mytilene (Greece), September 2007.]}

\end{abstract}
\newpage
\setcounter{page}{1}
\renewcommand{\theequation}{\arabic{section}.\arabic{equation}}
\section{Introduction}\label{introduction}
\setcounter{equation}{0}

The black hole information paradox is probably the most important issue for fundamental physics today. If we cannot understand its resolution, then we cannot understand how quantum theory and gravity work together. Yet very few people seem to understand how robust the original Hawking arguments are, and what exactly it would take to resolve the problem. 

In this review I try to explain the power of this paradox using mostly pictures. In section \ref{theorem}, I formulate the paradox as a `theorem': if quantum gravity effects are confined to within the planck length and the vacuum is unique, then there {\it will} be information loss. I conclude with a brief outline of how the paradox is resolved in string theory: quantum gravity effects are not confined to a bounded length (due to an effect termed `fractionation'), and the information of the hole is spread throughtout its interior, creating a `fuzzball'.

\section{Puzzles with black holes}

There are two closely connected problems that arise when we consider quantum theory in the context of black holes: the `entropy puzzle' and the `information paradox'.

\subsection{The entropy puzzle}

Take a box containing some gas, and throw it into a black hole. The gas had some entropy, so after the gas has vanished into the singularity, we have decreased the entropy of the Universe, and violated the second law!

Of course this sounds silly: if we threw the box into a trash can, then its entropy would be inside the trash can, whether we wanted to look in there or not. The black hole case  is a little different however,  since it is not clear how we would look into the hole to see the entropy of the gas. Nevertheless, physical intuition tells us that the entropy of the hole should have gone up when it swallowed the box of gas.

When the box falls into the hole, it increases the mass of the hole, and therefore the size of its horizon. 
Careful work with thermodynamics shows that we should attribute a `Bekenstein  entropy' \cite{matbek}
\begin{equation}
S_{bek}={A\over 4G}
\label{matone}
\end{equation}
to the black hole, where $A$ is the area of the horizon, and we have set $c=\hbar=1$. Then we have
\begin{equation}
{dS_{total}\over dt}={dS_{matter}\over dt}+{dS_{bek}\over dt}\ge 0
\end{equation}
and the second law of thermodynamics is saved.

This looks nice, but thinking a bit more, we find a deeper puzzle. From statistical physics we know that the entropy of any system is given by $S=\ln {\cal N}$, where ${\cal N}$ is the number of states of the system for the given macroscopic parameters. Applying this to the black hole, we should find
\begin{equation}
{\cal N}=e^{S_{bek}} 
\label{matthree}
\end{equation}
states for a black hole of given mass. Note that (\ref{matone}) is the area of the horizon measured in planck units. Thus for a solar mass black hole with a horizon radius $\sim 3 ~ Km$, we would have
\begin{equation}
{\cal N}\sim 10^{10^{77}}
\label{mattwo}
\end{equation}
states, an enormous number! Where should we look for these microstates? Since $S_{bek}$ is proportional to the horizon area, people tried to look for small `deformation modes' of the horizon. But it turns out that there is no such deformation in general; any excitation near the horizon either falls  to the singularity or flows off to infinity, leaving  a spherically symmetric horizon again. This observation came to be called `black holes have no hair', signifying that the horizon cannot hold any information in its vicinity. But if we find a unique geometry for the hole then the entropy would be $S=\ln 1=0$, in sharp contrast to (\ref{mattwo}). 

One may therefore think that the entropy is somehow at the singularity; after all the matter that made the hole in the first place disappeared into this singularity. In that case we would not see the differences between microstates in the classical geometry of the hole, and quantum effects at the singularity would differentiate the different states. But as we will see now, this possibility leads to an even more serious problem: the information paradox.

\subsection{The information paradox}

We have seen above that a black hole has entropy $S_{bek}$. It has an energy $E=M$, where $M$ is the mass of the hole. One may therefore ask if we could have the usual thermodynamic relation
\begin{equation}
TdS_{bek}=dE
\end{equation}
This would imply that the black hole has a temperature
\begin{equation}
T=\left ( {dS\over dE}\right )^{-1}=\left ( {d\over dM} ({4\pi (2GM)^2\over 4G})\right )^{-1}
={1\over 8\pi GM}
\end{equation}
If the black hole has a temperature, should it radiate? Temperature by itself does not imply radiation, but by the law of detailed balance in thermodynamics what we {\it can} say is that if the black hole can absorb quanta of a certain wavenumber with cross section $\sigma(k)$, then it should radiate the same quanta at a rate
\begin{equation}
\Gamma=\int {d^3 k\over (2\pi)^3}\sigma(k){1\over e^{\omega(k)\over T}-1}
\label{matqthir}
\end{equation}
But we know that $\sigma(k)$ is nonzero, since quanta can fall into the hole. Thus we must get radiation from the hole.

But the classical geometry of the hole does not allow any worldlines to emerge from the horizon! How then will we get this radiation? The answer, discovered by Hawking \cite{mathawking}, is that we must consider quantum processes; more precisely, quantum fluctuations of the vacuum. In the vacuum pairs of particles and antiparticles are continuously being created and annihilated. Consider such fluctuations for electron-positron pairs. Suppose we  apply a strong electric field in a region which is pure vacuum. When an electron-positron pair is created, the electron gets pulled one way by the field and the positron gets pulled the other way. Thus instead of annihilation of the pair, we can get creation of real (instead of virtual) electrons and positrons which can be collected on opposite ends of the vacuum region. Thus we get a current flowing through the space even though there is no material medium filling the region where the electric field is applied. This is called the `Schwinger effect'. 

A similar effect happens with the black hole, with the effect of the electric field now replaced by the gravitational field.  We do not have particles that are charged in opposite ways under gravity. But the attraction of the black hole falls off with radius, so if one member of  a particle-antiparticle pair is just outside the horizon it can flow off to infinity, while if the other member of the pair is just inside the horizon then it can get sucked into the hole. The particles flowing off to infinity represent  the `Hawking radiation' coming out of the black hole. Doing a detailed computation, one finds that the rate of this radiation is given by (\ref{matqthir}).
Thus we seem to have a very nice thermodynamical physics of the black hole. The hole has entropy, energy, temperature, and radiates as a thermal body should. 

But there is a deep problem arising out of the {\it way} in which this radiation is created by the black hole. As we will discuss in detail in the coming sections, the radiation which emerges from the hole is not in a `pure quantum state'. Instead, the emitted quanta are in a `mixed state' with excitations which stay inside the hole. There is nothing wrong in this by itself, but the problem comes at the next step. The hole loses mass because of the radiation, and eventually disappears. Then the quanta in the radiation outside the hole are left in a state that is `mixed', but we cannot see anything anything that they are mixed with! Thus the state of the system has become a `mixed' state in a fundamental way. This does not happen in usual quantum mechanics, where we start with a pure state $|\Psi\rangle$, and evolve it by some Hamiltonian $H$ as $|\Psi'\rangle=e^{-iHt}|\Psi\rangle$ to get another pure state at the end. We will describe mixed states in detail later, but for now we note that mixed states arise in usual physics when we coarse-grain over some variables and thereby discard some information about a system. This coarse-graining is done for convenience, so that we can extract the gross behavior of a system without keeping all its fine details, and is a standard procedure in statistical mechanics. But  there is always a `fine-grained' description available with all information about the state, so that underlying the full system there is always a pure state. With black holes we seem to be getting a loss of information in a fundamental way. We are not throwing away information for convenience; rather we cannot get a pure state even if we wanted. This implies a fundamental change in  quantum theory, and Hawking advocated that in the presence of gravity (which will make black holes) we should not formulate quantum mechanics with pure states and unitary evolution operators. Rather, we should think of mixed states as being basic, and describe these in terms of their `density matrices'. The evolution of these density matrices will be given not by the $S$ matrix but by  a dollar matrix $\$$ \cite{mathawking}. 

This was a radical proposal, and most physicists were not happy to abandon ordinary quantum mechanics when it works so well in all other contexts. But if we are to bypass this `information paradox' then we have to see how exactly how this radiation is emitted and what changes to the physics could make this radiation emerge in a pure state.  Enormous effort has been spent on this problem. With string theory, we will see that we can now obtain a resolution of the paradox. Perhaps it should not be surprising that this resolution itself comes with a  radical change in our understanding of how quantum gravity works. Earlier attempts at resolving the paradox had assumed that quantum gravity effects operate over distances of order the planck length or less. This seems natural since the only fundamental length scale that we can make out of the fundamental constants $c, \hbar, G$ is
\begin{equation}
l_p=\left ( {\hbar G\over c^3}\right )^{1\over 2}\sim 10^{-33} ~ {cm}
\end{equation}
As we will see below, if it were indeed true that all quantum gravity effects were confined to within a length scale like $l_p$ (or any other fixed length scale) then we would get information loss, and quantum mechanics would need to be changed. But how can we get any other natural length scale for quantum gravity effects? If we collide two gravitons then it is true that quantum gravity effects should start when the wavelengths of the gravitons become order $l_p$. But a black hole is made up of a large number of quanta $N$; the larger the black hole the larger this number $N$. Then we have to ask if quantum gravity effects extend over distances $\sim l_p$ or over distances $N^\alpha l_p$ where $\alpha$ is some appropriate constant. In string theory we find that the latter is true, and that $N, \alpha$ are such that the length scale of quantum gravity effects becomes of the order of the radius of the horizon. This changes the process by which the radiation is emitted, and the radiation can emerge in a pure state. 

\subsection{The plan of the review}

There exist many  reviews on the subject of black holes, and there are also reviews of the `fuzzball' structure emerging from string theory \cite{matfuzz}.  
What I will do here is  a bit different: I will try to give a detailed pictorial description of the information problem. We will study the black hole geometry in detail, and see how wavemodes evolve to create Hawking radiation. Then we will discuss the `mixed' nature of the quantum state that is created in this radiation process. Most importantly, we will discuss  why the argument of Hawking showing information loss is {\it robust}, and can only be bypassed by a radical change in one of the fundamental assumptions that we usually make about quantum gravity. We will close with a brief summary of black holes in string theory and the fuzzball nature of the black hole interior.

\section{Particle creation in curved space}
\label{matsec:2}

The story of Hawking radiation really begins with the understanding of particle creation in curved spacetime. (For reviews see \cite{matbirrel}.) Particles are described in terms of an underlying quantum field,  say a scalar field $\phi$. We can write a covariant action for this field, and do a path integral. But how do we define particles? In flat space we expand the field operator as
\begin{equation}
\hat\phi=\sum_{\vec k}{1\over \sqrt{V}}{1\over \sqrt{2\omega}}\left (\hat a_{\vec k} e^{i\vec k\cdot \vec x-i\omega t}+\hat a ^\dagger _{\vec k}e^{-i\vec k\cdot \vec x +i\omega t}\right )
\end{equation}
where $V$ is the volume of the spatial box where we have taken the field to live, and $\omega=\sqrt{|\vec k|^2+m^2}$ for a field with mass $m$. The vacuum is the state annihilated by all the $\hat a$
\begin{equation}
\hat a_{\vec k}|0\rangle=0
\end{equation}
and the $\hat a^\dagger_{\vec k}$ create particles.

In {\it curved} spacetime, on the other hand, there is no canonical definition of  particles. We can choose any coordinate $t$ for time, and decompose the field into positive and negative frequency modes with respect to this time $t$. Let the positive frequency  modes be called $f(x)$; then their complex conjugates give negative frequency modes $f^*(x)$. The field operator can be expanded as
\begin{equation}
\hat\phi(x)=\sum_n \left (\hat a_n f_n(x)+\hat a^\dagger_n f_n^*(x)\right  )
\end{equation}
Then we can define a vacuum state as one that is annihilated by all the annihilation operators
\begin{equation}
\hat a_n|0\rangle_a=0
\end{equation}
The creation operators generate particles; for example a 1-particle state would be
\begin{equation}
|\psi\rangle=\hat a_n^\dagger |0\rangle_a
\end{equation}
We have added the subscript $a$ to the vacuum state to indicate that the vacuum is defined with respect to the operators $\hat a_n$. But since there is no unique choice of the time coordinate $t$, we can choose a different one $\tilde t$. We will then have a different set of positive and negative frequency modes, and an expansion
\begin{equation}
\hat\phi(x)=\sum_n \left (\hat b_n h_n(x)+\hat b^\dagger_n h_n^*(x)\right  )
\end{equation}
Now the vacuum would be defined as
\begin{equation}
\hat b_n |0\rangle_b=0
\end{equation}
and the $\hat b^\dagger_n$ would create particles. 

The main point now  is that a person using the operators $\hat a, \hat a^\dagger$ would think that $|0\rangle_a$ was a vacuum, but he would not think that the state $|0\rangle_b$ was a vacuum -- he would find it to contain particles of the type created by the $\hat a^\dagger_n$. Let us see how one finds exactly how many $\hat a^\dagger$ particles there are in the state $|0\rangle_b$. The mode functions $f_n$ are normalized using an inner product defined as follows. Take any spacelike hypersurface, with volume element $d\Sigma^\mu$ (thus the vector $d\Sigma^\mu$ points normal to the hypersurface and has a value equal to the volume of the surface element). Then
\begin{equation}
(f, g)\equiv -i\int d\Sigma^\mu \left ( f\partial_\mu g^*-g^*\partial_\mu f \right )
\end{equation}
Under this inner product we will have
\begin{equation}
(f_m, f_n)=\delta_{mn}, ~~~(f_m, f^*_n)=0, ~~~(f^*_m, f^*_n)=-\delta_{mn}
\end{equation}
Now from the two different expansions of $\hat \phi$ we have
\begin{equation}
\sum_n \left (\hat a_n f_n(x)+\hat a^\dagger_n f_n^*(x)\right  )
=\sum_n \left (\hat b_n h_n(x)+\hat b^\dagger_n h_n^*(x)\right  )
\end{equation}
Taking the inner product with $f_m$ on each side, we get
\begin{equation}
\hat a_m=\sum_n(h_n, f_m)\hat b_n + \sum_n(h^*_n, f_m) \hat b_n^\dagger\equiv \sum_n \alpha_{mn} \hat b_n+\sum_n \beta_{mn} \hat b^\dagger_n
\end{equation}
Thus the vacuum $|0\rangle_a$ satisfies
\begin{equation}
0=\hat a_m|0\rangle_a=\left (\sum_n \alpha_{mn} \hat b_n+\sum_n \beta_{mn} \hat b^\dagger_n\right )|0\rangle_a
\label{matqtwo}
\end{equation}
Let us see how to solve this equation. Suppose we had just one mode, with a relation
\begin{equation}
(b+\gamma b^\dagger)|0\rangle_a=0
\label{matqone}
\end{equation}
The solution to this equation is of the form
\begin{equation}
|0\rangle_a=Ce^{\mu\hat b^\dagger \hat b^\dagger}|0\rangle_b
\end{equation}
where $C$ is a normalization constant and $\mu$ is a number that we have to determine. Expand the exponential in a power series
\begin{equation}
e^{\mu\hat b^\dagger \hat b^\dagger}=\sum_n {\mu^n\over n!} (\hat b^\dagger \hat b^\dagger)^n
\end{equation}
With a little effort using the commutator $[\hat b, \hat b^\dagger]=1$ we find that
\begin{equation}
\hat b  (\hat b^\dagger \hat b^\dagger)^n=(\hat b^\dagger \hat b^\dagger)^n\hat b + 2n\hat b^\dagger (\hat b^\dagger \hat b^\dagger)^{n-1}
\end{equation}
Putting this in the series for the exponential, we find that
\begin{equation}
\hat b e^{\mu\hat b^\dagger \hat b^\dagger}|0\rangle_b=2\mu \hat b^\dagger e^{\mu\hat b^\dagger \hat b^\dagger}|0\rangle_b
\end{equation}
Looking at (\ref{matqone}) we see that we should choose $\mu=-{\gamma\over 2}$, and we get
\begin{equation}
|0\rangle_a=C e^{-{\gamma\over 2}\hat b^\dagger \hat b^\dagger} |0\rangle_b
\label{matqfour}
\end{equation}
This state has the form
\begin{equation}
|0\rangle_a=C|0\rangle_b+C_2\hat b^\dagger \hat b^\dagger|0\rangle_b+C_4 
\hat b^\dagger \hat b^\dagger\hat b^\dagger \hat b^\dagger|0\rangle_b+\dots
\label{matqthree}
\end{equation}
so it  looks like a part that is the $ b$ vacuum, a part that has two particles of type $b$, a part with four such particles, and so on.

Returning to our full equation (\ref{matqtwo}) we have the solution
\begin{equation}
|0\rangle_a=C e^{-{1\over 2}\sum_{m,n}\hat b_m^\dagger\gamma_{mn} \hat b^\dagger_n} |0\rangle_b
\label{matdone}
\end{equation}
where the matrix $\gamma$ is symmetric and is given by 
\begin{equation}
\gamma={1\over 2} \left ( \alpha^{-1} \beta+(\alpha^{-1} \beta)^T\right )
\end{equation}

To summarize, there are many ways to define time and therefore many ways to define the vacuum and particles in curved space. The vacuum in one definition looks, in general, full of particle pairs in other definitions. How then are we going to do any physics with these particles?

What helps is that we will usually detect particles in some region which is far away from the region where spacetime is curved, for example at asymptotic infinity in a black hole geometry. There is a natural choice of coordinates at infinity, in which the metric looks like $\eta_{\mu\nu}$. We can still make boosts that keep the metric in this form, but the change of  time coordinate under these boosts does not change the vacuum. What happens is that positive frequency modes change to other positive frequency modes, giving the expected change of the energy of a quantum when it is viewed from a moving frame. 

But even though this may be a natural choice of coordinates, giving a natural definition of particles, we may still ask why we cannot use some other other curvilinear coordinate system and its corresponding particles. The point is that we have to know the following physics at some point: what is the energy carried by these particles? This information is not given by the definition of the particle modes; rather, we need to know the energy-momentum tensor for these particle states. For the physical fields that we consider, we assume that the particles defined in the flat coordinate system with metric $\eta_{\mu\nu}$ are the ones which give the expected physical energy of the state, an energy which shows up for example in the gravitational attraction between these particles. 

So there is no ambiguity in how particles are defined at infinity, but if there is some region of spacetime which is curved, then wavemodes that  travel through that region and back out to spatial infinity can have a nontrivial number of particles at the end, even though they may have started with no particles excited in them at the start. What we need now is to get some physical feeling for the length and time scales involved in this process of particle creation.

\subsection{Particle creation: physical picture}

\begin{figure}[ht]
\includegraphics[scale=.20]{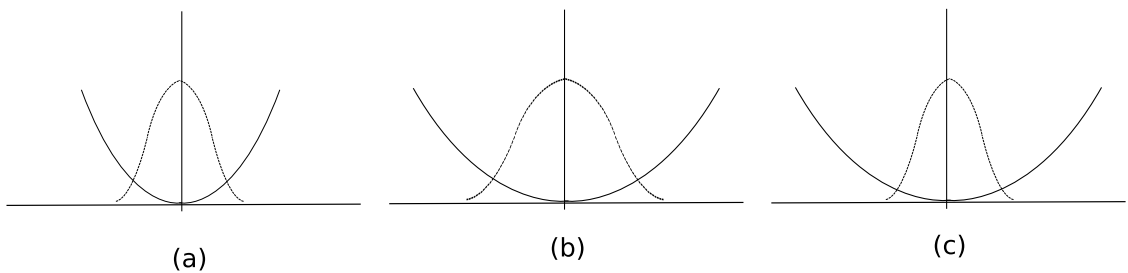}
%
%
\caption{(a) The potential characterizing a given fourier mode, and the vacuum wavefunction for this potential. (b) If the spacetime distorts, the potential changes to a new one, with its own vacuum wavefunction. (c) If the potential changes suddenly, we have the new potential but the old wavefunction, which will not be the vacuum wavefunction for this changed potential; thus we will see particles.}
\label{matfone}       
\end{figure}

Let us first get a simpler picture of why particles can get created when spacetime is curved. We know that each fourier mode of a quantum field behaves like a harmonic oscillator, and if we are in the excited state $|n\rangle$ for this oscillator then we have $n$ particles in this fourier mode. Thus the amplitude of this fourier mode, which we call $a$, has a Lagrangian of the form
\begin{equation}
L={1\over 2}\dot a^2- {1\over 2}\omega^2 a^2
\end{equation}
But as we move to later times, the spacetime can distort, and the frequency of the mode can change, so that we will get
\begin{equation}
L'= {1\over 2}\dot a^2- {1\over 2}\omega'^2 a^2
\end{equation}
We picture this situation in fig.\ref{matfone}. Fig.\ref{matfone}(a) shows the potential where the frequency is $\omega$. Let us require that here are no particles present in this fourier mode. Then we will have the vacuum wavefunction $|0\rangle$ for this harmonic oscillator. Now suppose we  change the potential to the one for frequency $\omega'$; this potential in shown in fig.\ref{matfone}(b). For this new potential, the vacuum state is a different wavefunction from the one for frequency $\omega$, and we sketch it in fig.\ref{matfone}(b). 

First suppose that the change of  frequency from $\omega$ to $\omega'$ was very slow. Then we will find that the vacuum wavefunction  will keep changing as the potential changes, in such a way that it remains the vacuum state for whatever potential we have at any given time. In particular when we reach the final potential with frequency $\omega'$, the vacuum wavefunction of fig.\ref{matfone}(a) will have become the vacuum wavefunction of fig.\ref{matfone}(b). This fact follows from the `adiabatic theorem', which describes the evolution of states when the potential changes slowly.

Now consider the opposite limit, where the potential changes from the one in fig.\ref{matfone}(a) to the one in fig.\ref{matfone}(b) very {\it quickly}. Then the wavefunction has had hardly any time to evolve, and we get the situation in fig.\ref{matfone}(c). The potential is that for frequency $\omega'$, but the wavefunction is still the vacuum wavefunction for frequency $\omega$. This is not the vacuum wavefunction for frequency $\omega'$, but we can expand it in terms of the wavefunctions $|n\rangle_{\omega'}$ which describe the level $n$ excitation of the harmonic oscillator for frequency $\omega'$
\begin{equation}
|0\rangle_\omega=c_0|0\rangle_{\omega'}+c_1 |1\rangle_{\omega'}
+c_2|2\rangle_{\omega'}+\dots
\end{equation}
Actually since the wavefunction that we have is symmetric under reflections $a\rightarrow -a$, we will get only the even levels $|n\rangle$ in our expansion
\begin{equation}
|0\rangle_\omega=c_0|0\rangle_{\omega'}+c_2|2\rangle_{\omega'}+c_4|4\rangle_{\omega'}+\dots
\end{equation}
This is like the expansion (\ref{matqthree}), and a little more effort shows that the coefficients $c_n$ will be of the form that will give the 
exponential form (\ref{matqfour}). 

Thus under slow changes of the potential the fourier mode remains in a vacuum state, while if the changes are fast then the fourier mode gets populated by particle pairs. But what is the timescale that distinguishes slow changes from fast ones? The only natural timescale in the problem is the one given by frequency of the oscillator
\begin{equation}
\Delta T\sim \omega^{-1}\sim \omega'^{-1}
\end{equation}
where we have assumed that the two frequencies involved are of the same order. If the potential changes over times that are {\it small} compared to $\Delta T$, then in general  particle pairs will be produced.

We can now put this discussion in the context of  curved spacetime. Let the variations of the metric be characterized by the length scale $L$; i.e., the length scale for variations of $g_{ab}$ is $\sim L$ in the space and time directions, and the region under consideration also has length $\sim L$ in the space and time directions. We assume that the metric  varies significantly (i.e. $\delta g\sim g$) in this region. Then the particles produced in this region will have a wavelength $\sim L$ and the number of produced particles will be order unity. Thus there is no other `large dimensionless number' appearing in the physics, and the length scale $L$ governs the qualitative features of  particle production. 

An example of such a metric variation would be if we take a star with radius $6GM$ (so it is not close to being a black hole), and then this star shrinks to a size $4GM$
(still not close to a black hole) over a time of order $\sim GM$. Then in this process we would produce order unity number of quanta for the scalar field, and these quanta will have  wavelengths $\sim GM$. After the star settles down to its new size, the metric becomes time independent again, and their is no further particle production.

As it stands, this particle production is a very small effect, from the point of view of energetics. In the above example, the length $GM$ is of order kilometers or more, so  the few quanta we produce will have wavelengths  of the order of kilometers. The energy of these quanta will be very small, much smaller than the energy  $M$ present  in the star which created the changing metric. So particle production can be ignored in most cases where the metric is changing on astrophysical length scales.

We will see that a quite different situation emerges for the black hole, where particle production keeps going on until all the mass of the black hole is exhausted.

\subsection{Particle production in black holes}

\begin{figure}[ht]
\includegraphics[scale=.25]{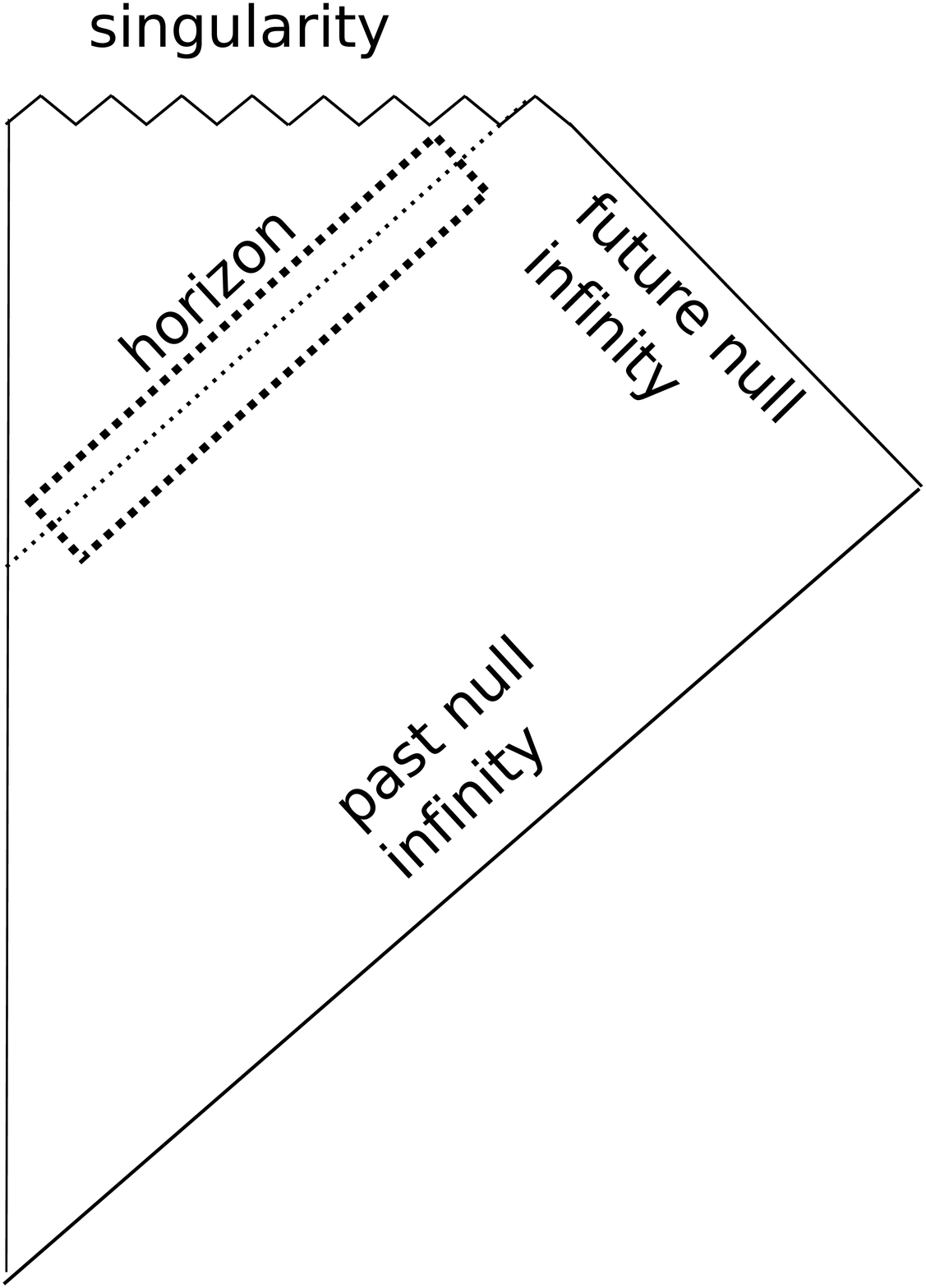}
%
%
\caption{The Penrose diagram for a black hole (without the backreaction effects of Hawking evaporation). Null rays are straight lines at $45^o$. Thus we see that the horizon is a null surface. Hawking radiation collects at future null infinity. }
\label{matfsix}       
\end{figure}

The metric of a Schwarzschild hole is often written as
\begin{equation}
ds^2=-(1-{2GM\over r})dt^2+{dr^2\over 1-{2GM\over r}}+r^2 (d\theta^2+\sin^2\theta d\phi^2)
\label{matqsix}
\end{equation}
This metric looks time independent, so we might think at first that there should be {\it no} particle production. If we had a time independent geometry for a star, there would indeed be no particle production. What is different in the black hole case? The point is that the coordinate system in the above metric covers only a part of the spacetime -- the part outside the horizon $r=2GM$. Once we look at the full metric we will not see a time independent geometry. The full geometry is traditionally described by a Penrose diagram, which we sketch in fig.\ref{matfsix}. The region of this diagram where the particle production will take place is indicated by the box with dotted outline around the horizon. 

\begin{figure}[ht]
\includegraphics[scale=.25]{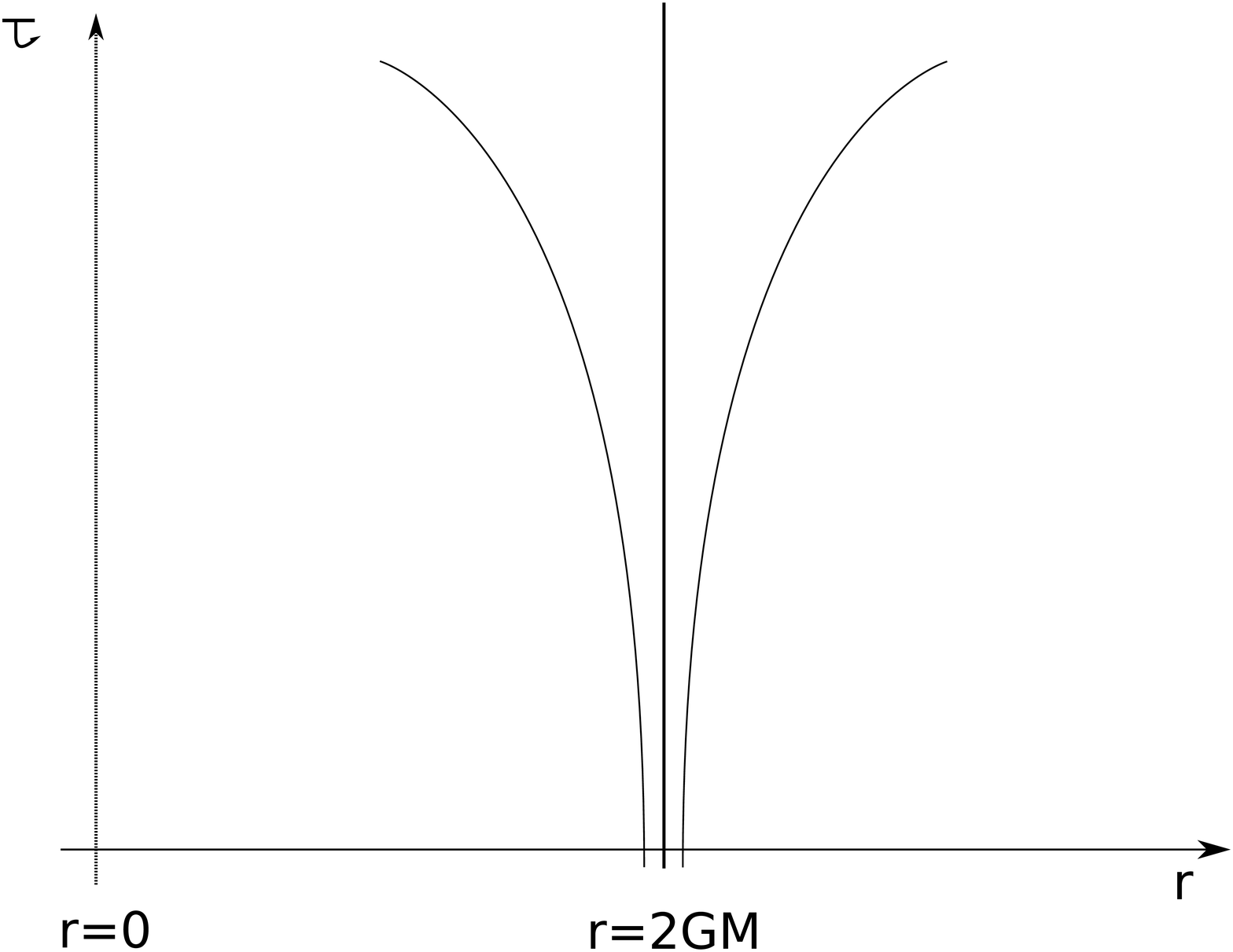}
%
%
\caption{A schematic picture of the dotted box in fig.\ref{matfsix}. The horizon has been rotated to be vertical. One coordinate is $r$. The other axis has been called $\tau$, but there is no canonical choice of $\tau$ (the metric will degenerate at the horizon anyway if we try to make it independent of $\tau$). We see that the null geodesics on the two sides of the horizon move away from $r=2GM$ as they evolve.}
\label{matftwo}       
\end{figure}

From the Penrose diagram we can easily see which point is in the causal future of which point, but since lengths have been `conformally scaled' we cannot get a good idea of relative lengths at different locations on the diagram. Thus in fig.\ref{matftwo} we make a schematic sketch of the boxed region in fig.\ref{matfsix}. The horizontal axis is $r$, which is a very geometric variable in the problem -- the value of $r$ at any radius is given by writing the area of the 2-sphere at that point as $4\pi r^2$. The line at $r=0$ is the `center' of the black hole; thus this is a region of high curvature (the singularity)  after the black hole forms. The line $r=2GM$ is the horizon. Spatial infinity is on the right, at $r\rightarrow\infty$.

The vertical axis in fig.\ref{matftwo} is called $\tau$; it is some time coordinate that we have introduced to complement $r$. At large $r$ we let $\tau\rightarrow t$, where $t$ is the Schwarzschild time. The metric will {\it not} be good everywhere in the coordinates $(r,\tau)$; it will degenerate at the horizon. This will not matter since all  we want to do with the help of this figure is show how geodesics near the horizon evolve to smaller or larger $r$ values.

A massless particle that is at the horizon and trying its best to fly out never manages to escape, but stays on the horizon. This can be seen as follows. The massless particle follows a null geodesic. Let us allow no angular part to its momentum to ensure that all the momentum is directed radially outwards in the attempt to escape. Thus from the metric (\ref{matqsix}) we will have
\begin{equation}
0=ds^2=-(1-{2GM\over r})dt^2+{dr^2\over 1-{2GM\over r}}
\end{equation}
which gives
\begin{equation}
dr=(1-{2GM\over r}) dt
\end{equation}
So if we are on the horizon $r=2GM$ then we get $dr=0$, i.e. the particle stays on the horizon. 

What if the particle started slightly outside the horizon, and tried to fly radially outwards? Now it can escape, so after some time the particle will reach out to a larger radius, say $r\sim 3GM$. This null geodesic starts out near the horizon, but `peels off' towards infinity. 

Similarly, consider a massless particle that starts a little {\it inside} the horizon and tries to fly radially outwards. This time it cannot escape the hole or even remain where it started; this null geodesic `peels off' and falls in towards smaller $r$. The figure shows the geodesic reaching the radius $r\sim GM$ which is inside the hole, though still comfortably away from the singularity.

Now we see that in this vicinity of the horizon, there is a `stretching' of spacetime going on. A small region near the horizon gets `pulled apart' with the part inside the horizon moving deeper in, and the part outside the horizon moving out. We will make this more precise later, but we can now see that the metric indeed has a time dependence which can cause particle creation. Moreover, this stretching goes on as long as the black hole lasts, since whenever we have the horizon we will see such a `peeling off' of null geodesics from the two sides of the horizon. Thus if there will be particle production from this stretching of spacetime, it will keep going on till the black hole disappears and there is no more horizon.

\section{Slicing the black hole geometry}\label{matsecslice}

We have seen that the Schwarzschild coordinates cover only the exterior of the horizon, and so do not give a useful description of the spacetime for the purposes of understanding Hawking radiation. What we need is a set of spacelike slices that `foliate' the spacetime geometry, covering both the outside and the inside of the hole. Let us see how to make such spacelike slices.

\begin{figure}[ht]
\includegraphics[scale=.25]{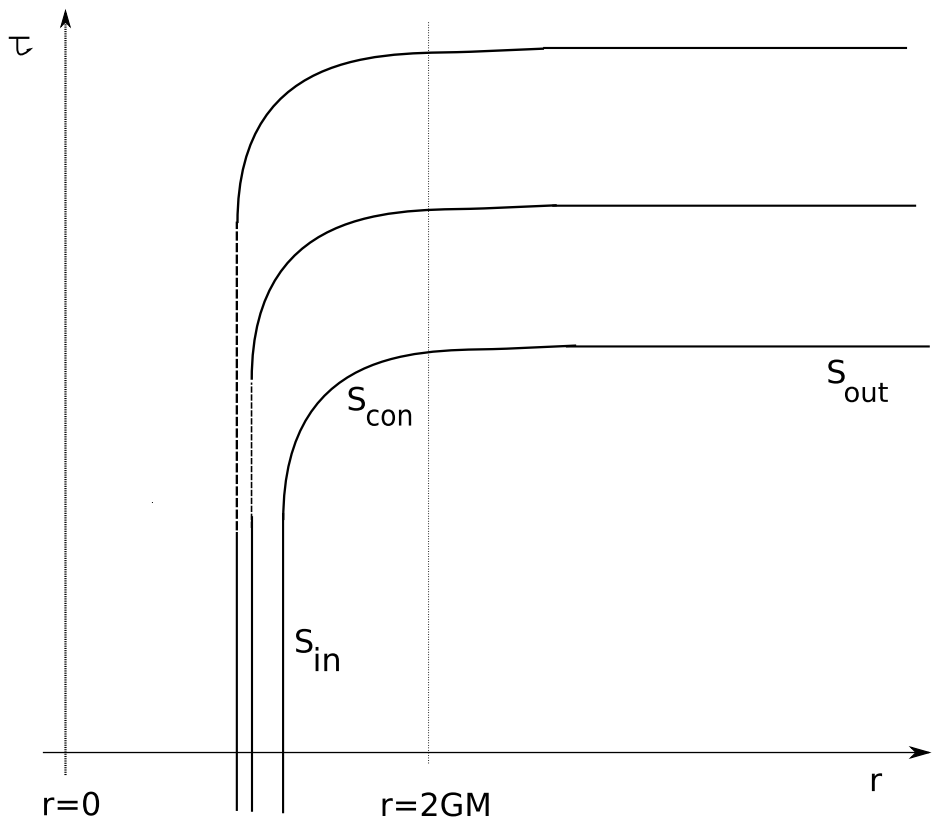}
%
%
\caption{Constructing a slicing of the black hole geometry. For $r>3GM$ we have the part $S_{out}$ as a $t=constant$ slice. The `connector' part $S_{con}$ is almost the same on all slices, and has a smooth intrinsic metric as the surface crosses the horizon. The inner part of the slice $S_{in}$ is a $r=constant$ surface, with the value of $r$ kept away from the singularity at $r=0$. The coordinate $\tau$ is only schematic; it will degenerate at the horizon. }
\label{matfthree}       
\end{figure}

Consider the slices sketched in fig.\ref{matfthree}. Far outside the horizon, we would like to have the spacelike slice look quite like a spacelike slice in ordinary flat spacetime. Thus we let it be the surface $t=constant$, all the way from infinity to say $r=3GM$, a point that is comfortably far away from the vicinity of the horizon. We call this part of the spacelike surface $S_{out}$.

What should we do inside the horizon? From the metric (\ref{matqsix})
we see that inside the horizon $r=2GM$ space and time interchange roles; i.e., the $t$ direction is spacelike while the $r$ direction is timelike. Thus for the part of the slice inside the horizon we use a $r=constant$ slice. Let us take this slice at $r=GM$, comfortably far away from the horizon at $r=2GM$ and also from the singularity at $r=0$. Let us call this part of the spacelike surface $S_{in}$.

We must now connect these two parts of our spacelike surface. It is not hard to convince oneself that this can be done with a smooth `connector' segment, which is everywhere spacelike. Let us call this segment of the spacelike surface $S_{con}$. 

One might be worried that this spacelike slice is not covering the region near $r=0$. Let us assume that the black hole formed at some time $t\sim t_{initial}$. Then for $t\ll t_{initial}$, there was no singularity at $r=0$. Thus imagine extending the part $S_{in}$ of the slice down to a time before this singularity, whereupon we bend it smoothly to reach $r=0$. 
(This part of the slice is not depicted in the figure, since it will not be of immediate use to us in the discussions that follow.)

All this makes one spacelike slice, but what we need is a family of such slices to foliate the region of spacetime that is of interest. Let us try to make a `later' slice in our foliation. For the part $S_{out}$  we know what this means: we should take $t=constant$ with a larger value of $t$. What do we do for the inside part $S_{in}$? If we wish to advance this part forward in the direction that is locally timelike, then we have to move it {\it inwards} towards smaller $r$ (recall that $r$ is the timelike direction inside the horizon). If we keep moving our successive slices towards smaller $r$, we will soon reach the vicinity of the singularity at $r=0$, which we did not want to do. So we will move each successive slice to a smaller $r$ value but only by a  very {\it small} amount; we will make this amount smaller and smaller so that the slices asymptote to say the surface $r={GM\over 2}$, still comfortably away from the singularity at $r=0$.

So what is the essential difference between one slice and a later slice? The outer part $S_{out}$ has just moved up in time $\tau$, but not changed its intrinsic geometry. The `connector' part $S_{con}$ has not changed its intrinsic geometry  much either. The nontrivial  change has been in the inner part $S_{in}$, which has not changed in its $r$ location very much, but it has become {\it longer}; there is an extra part indicated by the dotted segment that has emerged to allow $S_{in}$ to connect to the rest of the slice. 

\begin{figure}[ht]
\includegraphics[scale=.25]{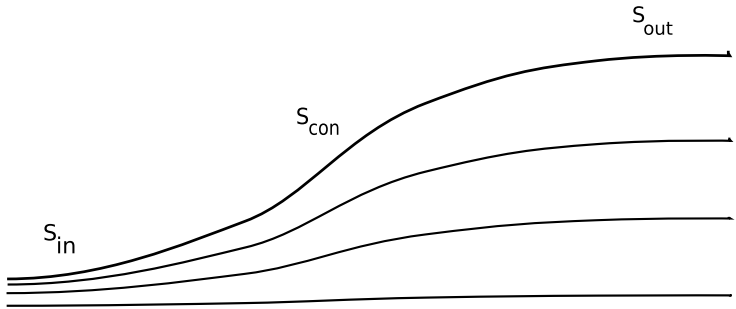}
%
%
\caption{The slices of fig.\ref{matfthree} redrawn in a different way to show the changes from one slice to the next.}
\label{matffour}       
\end{figure}

In the $r-\tau$ plane of fig.\ref{matfthree} this may look like a strange set of slices, so we redraw them a bit differently in fig.\ref{matffour}. The lowest slice corresponds to the time before the black hole is formed. Thus it is essentially a flat slice $t=constant$ all through. On later slices, the part on the right, which is in the `outer' region, keeps advancing forward in time. The part on the `inside' advances very little. As a consequence there is a lot of stretching in the part that connects the part on the left to the part on the right. Later and later slices have to stretch more and more in this region. 

In any spacetime we always have the freedom of pushing forward our spacelike slice at different rates at different locations (Wheeler terms this  `many-fingered' time in general relativity). But note that in flat spacetime for example we could not have done what we see in fig.\ref{matffour}. Thus consider flat spacetime, and let the first slice be $t=constant$. Now for later slices we try to keep the left side of the slice fixed (or advancing very slightly) and we make the right side move up to later times. Then after a while we will find that the part of the slice joining these two parts is no longer spacelike; it will become null somewhere and then become timelike. Thus the kind of slices that we see in 
fig.\ref{matffour} is particular to the black hole geometry, and the infinite stretching that we see in these slices can be traced to the presence of a horizon. 

\begin{figure}[ht]
\includegraphics[scale=.25]{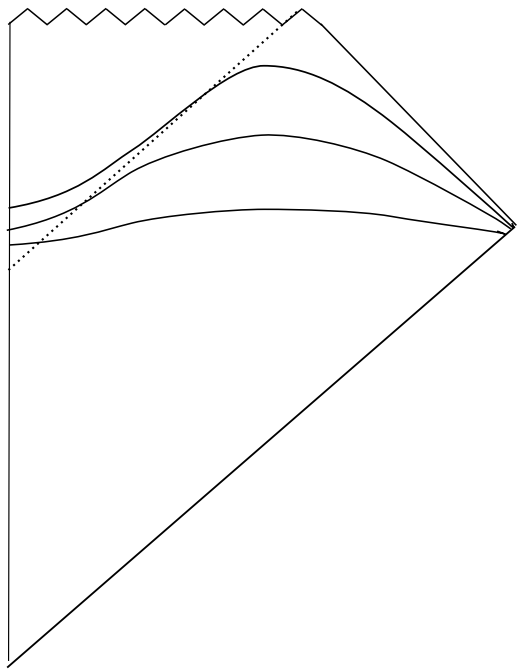}
%
%
\caption{The slices drawn on the Penrose diagram. Later slices go up higher near future null infinity and will thus capture more of the Hawking radiation.}
\label{matfseven}       
\end{figure}

Finally, in fig.\ref{matfseven} we depict the slices on the Penrose diagram. The later the time slice, the more it moves up near future null infinity before coming into the horizon. Thus the later and later time slices will be able to capture more and more  of the Hawking radiation emitted from the hole.

The important fact about slices in the black hole geometry is the following. In  Schwarzschild coordinates both $g_{tt}$ and $g_{rr}$ become singular at the horizon: one vanishes and one diverges. With the slices we have chosen, the spatial metric along the slices remains regular as we cross the horizon. If we allow our slices to reach the singularity at $r=0$, then we can foliate the geometry by slices which are spacelike, and which are all similar to each other as far as their intrinsic geometry is concerned. Then why will there be particle production? The point is that there is no {\it timelike killing vector in the geometry}. Suppose we draw a vector connecting a point at $r=r_0$ on one slice with the point at $r=r_0$ on the next slice. If this vector was timelike everywhere we could use it to define time evolution, and everything would look time independent: the slices don't change, and the metric with this choice of time direction will look time independent. But this vector will  {\it not} be timelike everywhere; it will become null on the horizon and be spacelike inside the horizon.

We have taken extra care to make our slices not approach the singularity -- we let them follow a $r=constant$ path to an early enough stage where the singularity had not formed, and then took them in to $r=0$. This feature of the slices is not directly related to the production of particles in Hawking radiation, but we have done the slicing in this way so that the evolution  stays in a domain where curvature is everywhere low and so classical gravity would appear to be trustworthy.

To summarize,  the central point that we we see with these different ways of exhibiting the slices is that the geometry of the black hole is not really a time independent one, and particle production can therefore be expected to happen.

\subsection{The wavemodes}

Let us now look at the wavemodes of the scalar field in the black hole geometry.

We will look at non-rotating holes, so the metric is spherically symmetric, and we can decompose the modes of the scalar field $\phi$ into spherical harmonics.  Most of the Hawking radiation turns out to be in the lowest harmonic, the s-wave, so we will just focus our attention on this $l=0$ mode; the physics extends in an identical way to the other harmonics. We will suppress the $\theta,\phi$ coordinates, drawing all waves only in the $r,t$ plane.

To study the emission from the hole, in principle we should solve the wave equation in the metric of the black hole. This is complicated, though many approximations have been developed to carry out this computation and a lot of numerical work has been done as well. But the basic ideas involved in the computation of Hawking radiation can be understood by using a very simple description of the wavemodes: solving them by the `eikonal approximation', which we now describe.

Since we have taken the harmonic $l=0$, all we have to do is describe the wave in the $r,t$ plane. In flat space, we  have two kinds of modes:  ingoing modes and outgoing modes. For higher $l$, there is a `centrifugal barrier' from angular momentum, and there will not be such a clean separation between ingoing and outgoing waves at small $r$. But if we are looking at a high frequency mode then this angular momentum term is ignorable, and the physics again splits into an ingoing and an outgoing mode. From the wave equation $\square \phi=0$ we find that these ingoing and outgoing modes travel at the speed of light. 

Consider an outgoing wavemode, and look at it on a spacelike hypersurface. Then we would see a sinosoidal oscillation of its phase with some wavelength which we call $\lambda$. We assume that
\begin{equation}
\lambda \ll GM
\end{equation}
Here $GM$ is the scale over which the metric of the black hole varies, so  we are asking that the wavelength be much smaller than the scale over which the metric is curved. Thus the wave oscillations will locally look like oscillations on a piece of flat spacetime. It will turn out that as our wavemodes evolve their wavelength will increase, and will finally become order $\sim GM$, but by the time that happens they would be waves traveling near infinity where we understand their physics very well. Thus while we may find the overall radiation rate to be incorrect by a factor of order unity because $\lambda\sim GM$ near the end of the evolution, 
the basic problem created by the `entanglement of Hawking pairs' will be very robust and will not be affected by the errors caused by our approximation.

\begin{figure}[ht]
\includegraphics[scale=.38]{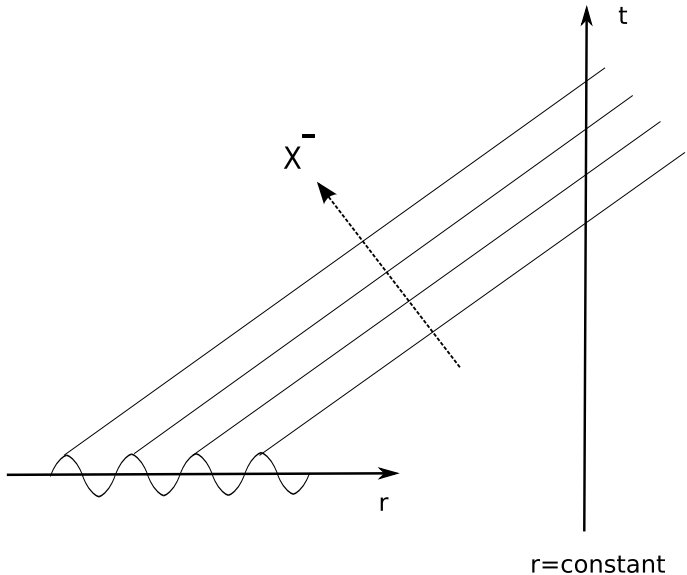}
%
%
\caption{The wavemode on an initial spacelike surface is evolved by letting the phase be constant on outgoing null rays. At infinity we can describe the mode by its intersection with a spacelike surface, which gives a function $\sim e^{ikr}$. Alternatively we can give its  intersection with a timelike surface which gives a function $\sim e^{-ik t}$. Lastly, we can describe the mode   by giving the phase on different outgoing null rays, which gives a function $e^{-ikX^-}$.}
\label{matffive}       
\end{figure}

In fig.\ref{matffive} we sketch  the wavemode as seen on a spacelike surface. At each point on the surface, the wavemode is a complex number given by an amplitude and a phase. 
Let this be an outgoing mode, of the type $e^{ik(r-t)}$ at infinity. Take a point $A$ on this spacelike surface,  and suppose the phase of the wavemode is $e^{i\phi_0}$ at this point. Draw a radial null geodesic through $A$, going out to infinity. Assign the phase $e^{i\phi_0}$ to all points on this null geodesic. Do the same for all points on the initial surface. The amplitude of the  wavemode at point $A$ also determines the amplitude at all points along the null geodesic through $A$, but we should note that in $d+1$ spacetime dimensions the amplitude of a spherical wave falls off as ${1\over r^{d-1\over 2}}$, so we put in this decrease with $r$ when finding the amplitude at all points along the null geodesic.

This process gives us a wavemode evolved to all points to the future of our spacelike hypersurface. If the wavelength was everywhere small compared to the curvature of the manifold, this would be a very good approximation to the actual solution of the wave equation; as it is, it will be an approximate solution that will serve our purpose in what follows. To summarize, we have evolved the outgoing wavemode by assuming that the phase of the mode stays constant along the outgoing null rays.

We can describe the wavemode in a few different ways. First, we can `catch' it on a spacelike surface as we did in fig.\ref{matffive}; in this case we see a waveform $e^{ikr}$ on the spacelike surface. We can also `catch' the wavemode by looking at its intersection with a timelike surface $r=const$. Then on this timelike surface we will see a phase like $e^{-i\omega t}$; this can be seen from the way the null lines intersect the timelike surface $r=constant$.  Thirdly we can describe the wavemode by giving the phase on each null ray. For outgoing waves the null rays are of the form 
\begin{equation}
t-r\equiv X^-=constant
\end{equation}
Thus we can write outgoing modes as
$e^{-ikX^-}$, or mode generally as $f(X^-)$. We will generally chose the function $f$ so as to make a localized wavepacket.

\section{The evolution of modes in the black hole}

In this section we will put together a lot of the tools we developed in the above discussion. Our goal is to look at wavemodes in the black hole background, and to see how they evolve. At the end of this evolution  the initial vacuum modes will be  populated with particles. What we wish to understand is the nature of this state with particles; in particular, how the various particles are correlated or `entangled' with each other. The entire essence of the information paradox lies in understanding this entanglement.

\begin{figure}[ht]
\includegraphics[scale=.25]{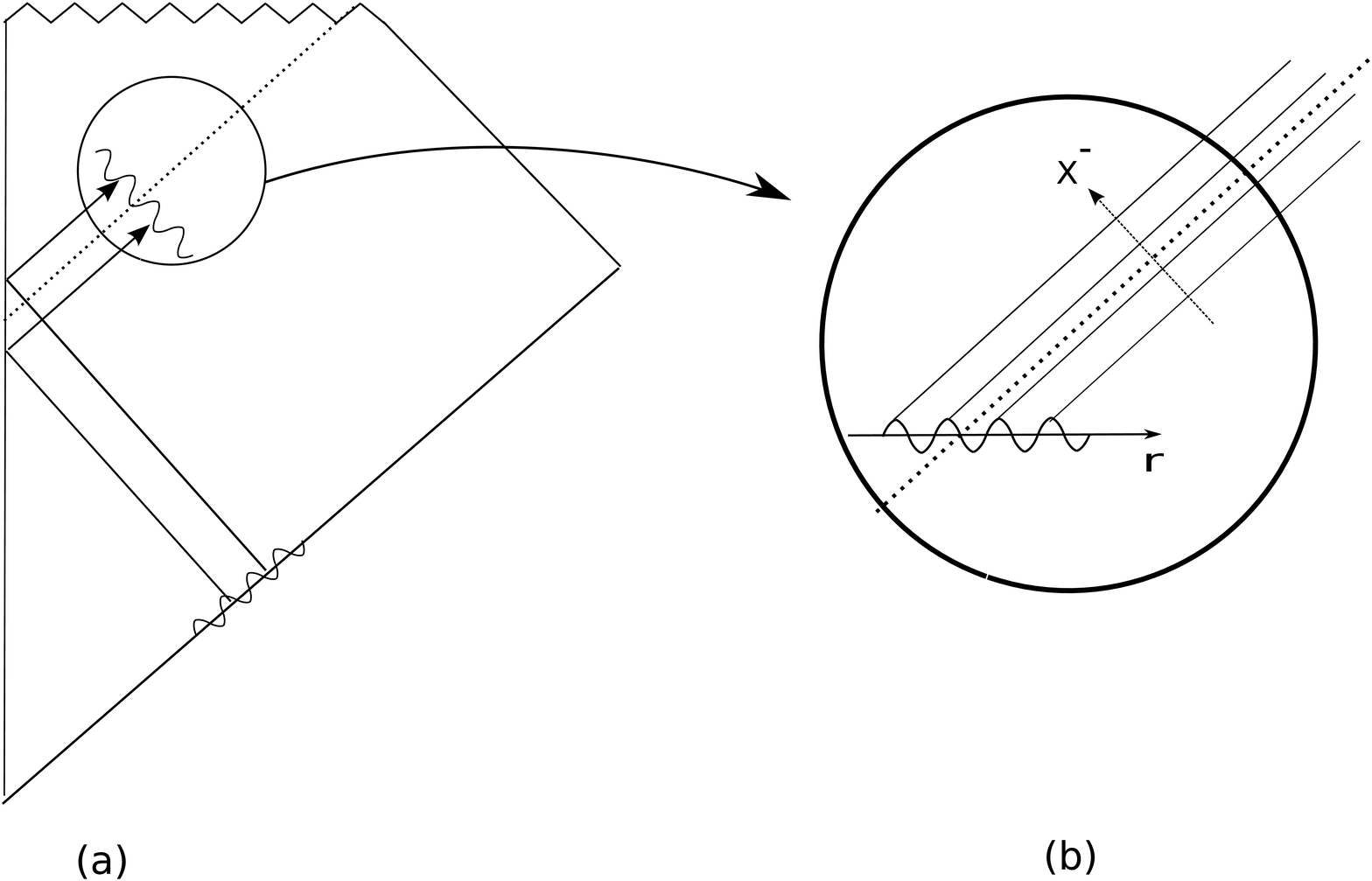}
%
%
\caption{(a) The region around the horizon is a vacuum. (b) An outgoing wavemode on an initial spacelike surface is evolved by letting the phase be constant on outgoing null geodescis.}
\label{matfeight}       
\end{figure}

Let us begin our discussion with a look at the Penrose diagram again, sketched in fig.\ref{matfeight}(a). We have drawn a circle around the region that is of immediate interest to us. 
In fig.\ref{matfeight}(b) we have drawn an expanded  view of this region.  The important thing about this region of spacetime is that in the traditional black hole picture this is a region of `empty space'. Thus there is no large curvature here or any other matter that our scalar field $\phi$ could interact with.  To understand better the state of the scalar field here, consider the evolution of field modes depicted in fig.\ref{matfeight}(a). Vacuum modes start off at past null infinity as ingoing modes. They reach $r=0$ and  scatter back as {\it outgoing} modes.  At this stage there is no singularity at $r=0$, so we just get outgoing vacuum modes after this scattering. These outgoing modes then show up in the circled region of fig.\ref{matfeight}(a). We are interested in the further evolution of these modes.

The outgoing field mode is drawn in more detail in  fig.\ref{matfeight}(b), where we have caught the mode on a spacelike surface which we will call our `initial slice'. 
We follow our above described method of evolving the wavemodes by letting the phase be constant along outgoing radial null geodesics. These null geodesics look like straight lines on the Penrose diagram, so at first it might seem that the wavelength of the mode is not changing as we follow the mode out towards infinity. This is not true, since in the Penrose diagram the actual distances between points is large when the points are near infinity. (In drawing the Penrose diagram we  squeeze the spacetime in a `conformal' way so that all of spacetime fits in a finite box; this automatically squeezes points near infinity by a large amount.)

\begin{figure}[ht]
\includegraphics[scale=.25]{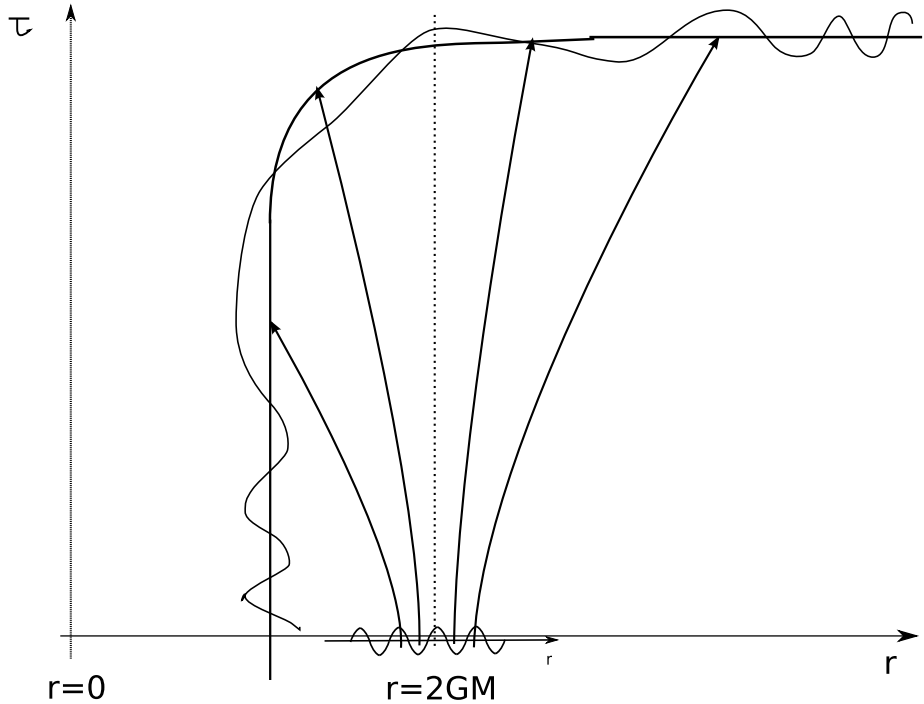}
%
%
\caption{A wavemode which is a positive frequency mode on the initial spacelike surface gets distorted when it evolves to a later spacelike surface;  the mode will not be made of purely positive frequencies after the distortion.}
\label{matften}       
\end{figure}

What we really want to see is how the wavelength of the mode changes as the mode is evolved. So in fig.\ref{matften} we sketch the evolution in the $r-\tau$ diagram that we discussed above. The initial slice is drawn again, with the outgoing wavemode on it. The lines of constant phase are drawn too, but now they do not look like straight lines. We had seen that the horizon itself is an outgoing null geodesic, that stays at all times at $r=2GM$. The rays starting slightly outside the horizon eventually `peel off' at go to spatial inifinity, while those starting slightly inside `peel off' and fall in towards small $r$. Thus the wavemode will get distorted as it evolves.

What we want to do now is to `catch' the wavemode on a later spacelike slice.  By following the null rays, we can obtain the phase of the wavemode all along this later slice. We can  see that there is quite a distortion between the wavemode as seen on the initial spacelike slice and the wavemode as it is `caught' on this later slice, and the changes come because the null geodesics just inside and just outside the horizon evolve in quite different ways.  But if the wavemode is distorted, there can be particle creation. We will now look at the distortion in much more detail, and discuss the nature of this particle creation.

\subsection{The coordinate map giving the expansion}

Consider the vicinity of the horizon sketched in fig.\ref{matfeight}(a). The local geometry is approximately flat space, and the field modes are in the vacuum state. Let us use null coordinates $y^+, y^-$ to describe the spacetime here (recall that the angular $S^2$ is suppressed throughout). The outgoing modes, which are of interest to us, are then of the form
\begin{equation}
\psi_{initial}\sim e^{ik y^-}
\end{equation}
We will assume that $k>0$. In the expansion of the field $\hat \phi$ the positive frequency modes multiply the annihilation operators $\hat a_k$. We will write the negative frequency modes as $e^{-iky^-}$; these will multiply the creation operators. 

Now let us see what coordinates would be good on the late time spacelike slice, sketched in fig.\ref{matften}. Consider the outer part of the slice $S_{out}$. This part is in a region which is close to flat Minkowski spacetime.We had discussed above that particles were well defined in such a region of spacetime, and this definition required us to use positive frequency modes based on the usual coordinates on Minkowski space.  So we  just use the standard definition of null coordinates here
\begin{equation}
X^+=t+r, ~~~X^-=t-r
\end{equation}
and the positive frequency modes are of the form
\begin{equation}
\psi_{out}\sim e^{iK X^-}
\end{equation}
with $K>0$. Since we evolve our field modes by keeping the phase of the mode constant on the outgoing null rays, all we need to know now is the relation  between the outgoing null coordinates on $S_{out}$ and the outgoing null coordinates on the initial slice
\begin{equation}
X^-=X^-(y^-)
\label{matqel}
\end{equation}
Note that $X^+$ is not involved in this relation, so modes that start off as functions of only $y^-$  become modes involving only $X^-$. 

What we need to know now is the nature of the function in (\ref{matqel}). This requires us to study the black hole metric and its geodesics. We will not carry out those computations here, but instead just quote the results and focus  on the qualitative physics which emerges. Detailed derivations of the results we will use can be found in \cite{mathawking,matwald,matparker}. A good review in the 2-d context can be found in \cite{matgiddings}.

First consider points very close to the horizon. Let $y^-=0$ be the horizon itself. Then points $y^-<0$ will be outside the horizon, and points $y^->0$ will be inside the horizon.

First consider null rays that are close to and just outside the horizon. 
It turns out that the null coordinate $X^-$ describing the rays at infinity is related to the label used near the horizon by a relation which is logarithmic
\begin{equation}
X^-\sim -\ln (-y^-)
\end{equation}
Note that $y^-<0$ for these rays, so we are taking the log of a positive number, as we should. Since $|y^-|$ is very small, the log is negative, so $X^-$ is actually positive. But as it stands this relation does not have the right units. The coordinate $y^-$ has units of length, so  we must first make a dimensionless variable and then take the log. The only natural length scale in the black hole geometry is $GM$, and the relation actually looks like
\begin{equation}
X^-=-(GM)~\ln \left ( - { y^-\over GM}\right )
\label{matqtw}
\end{equation}
This is a very interesting relation. A simple fourier mode $e^{iky^-}$ will get distorted by the logarithmic map. But to completely understand this logarithmic map we also need to understand what happens when the rays are {\it not} very close to the horizon. Thus look at  $y^- \lesssim -GM$. For such values of $y^-$ we are no longer close to the horizon. The rays are thus almost like rays in flat spacetime, and so there is no serious deformation of the wavemodes. Thus the relation 
(\ref{matqtw}) will change over to a relation like
\begin{equation}
X^-=-y^-, ~~~{\rm for}~~y^-\lesssim -GM
\end{equation}
and there will be {\it no} distortion of modes in this region which is away from the horizon.

Now let us look at the range $y^->0$, i.e., the part of the mode inside the horizon. 
There are no natural coordinates to describe the inside of the black hole. But since we have chosen a way of drawing spacelike slices across the entire geometry, we can use wavemodes here that are natural to this slicing; the actual choice of wavemodes inside the horizon will not matter at the end. Thus consider the part of the slice $S_{in}$ inside the horizon. We can introduce a coordinate $Y$ on this part of the slice which is linear in the distance measured along the slice. The null rays for $y^->0$ will intersect this slice at various points. We assign to each ray a null coordinate $Y^-$ which is equal to $Y$ at the point where the null ray intersects $S_{in}$.   Thus this coordinate assignment is similar to the coordinate $X^-$ defined on $S_{out}$, with the difference that in the case of $S_{out}$ there was a natural physical choice and particles defined using $X^-$ had the correct energy momentum tensor to be the real particles at infinity.

With such a coordinate choice $Y^-$, using the black hole geometry we find a relation like
\begin{equation}
Y^-\sim -\ln y^-
\end{equation}
or with the correct dimensionful parameters inserted
\begin{equation}
Y^-=-GM~\ln \left ( {y^-\over GM}\right )
\label{matqtwp}
\end{equation}
Thus there will be a distortion of the wavemodes on this side of the horizon as well.

The last point to note is that if $|y^-|$ is {\it very small} (i.e. the null ray is very close to the horizon) then the ray intersects the late time slice  on the `connector region' $S_{con}$, rather than on $S_{out}$ or $S_{in}$.

We now come to a crucial point. The  wavemode on the initial slice straddles both sides of  the horizon. Indeed, the horizon is not a `special place' in the geometry from a local point of view; this can be seen from the circle drawn in fig.\ref{matfeight}(a), which circles a region of spacetime much like any other. Thus wavemodes near the horizon naturally continue from one side of the horizon to the other. But the subsequent expansion of the geometry, encoded in the behavior of the null geodesics, treats the parts of the wavemode outside and inside the horizon quite differently. The part of the initial wavemode for $y^-\ll -GM$ does not `stretch', the part for $y^-$ negative and small in magnitude (but not too small) reaches $S_{out}$ with a logarithmic stretching, the part for {\it very} small $|y^-|$ ends up on the connector region $S_{con}$, and the part for $y^->0$ (but not too small)  ends up on $S_{in}$.  The consequent distortion of the wavemode is sketched in fig.\ref{matften}.

From this figure we can observe some basic facts about the distortion of the wavemode.   The distortion is large around the point where the rays move from being inside the horizon to being outside the horizon. As we will see in more detail below, there is very little distortion away from this region. Because the wavemode gets distorted, a given fourier mode on the initial slice becomes a combination of modes on the later slice. Note that all modes involved are outgoing modes; we have functions of $y^-, X^-, Y^-$. This fact is a consequence of our `ray approximation' where we evolve the mode by letting the phase of $\phi$ be constant on outgoing rays. 

The most important thing here is that the the part of the mode straddling the horizon splits into a part on $S_{out}$ and a part on $S_{in}$. This will make the state of the created particles a `mixed state' of the outside and inside quanta, as we shall discuss in more detail below.

\subsection{Detailed nature of the wavemode}

\begin{figure}[ht]
\includegraphics[scale=.20]{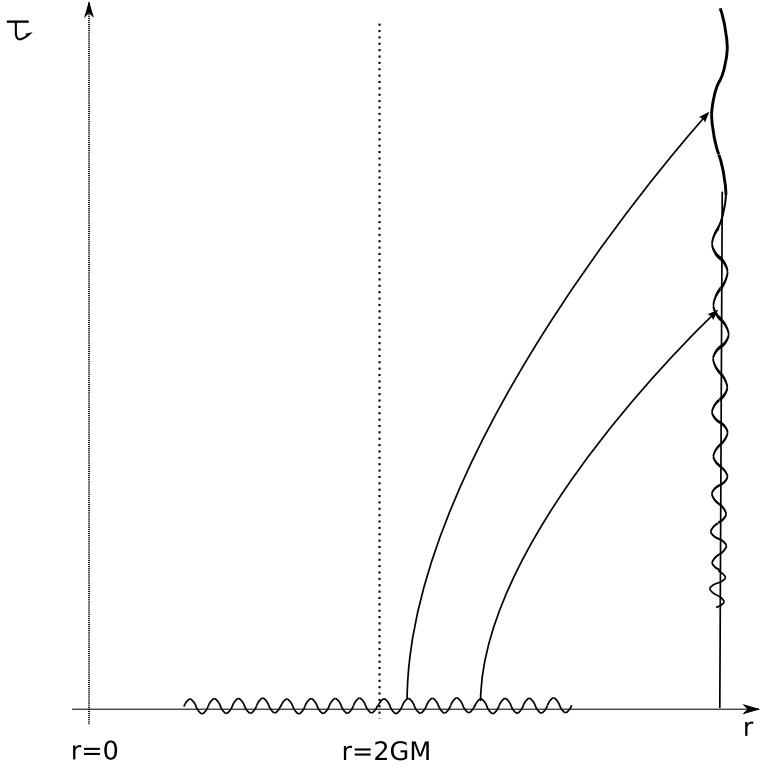}
%
%
\caption{The fourier mode on the initial spacelike slice is evolved in the eikonal approximation, and `caught' on the timelike surface $r=constant$ near infinity. (From the behavior of the mode on this surface we can immediately obtain what it looks like on any spacelike surface near infinity.) The wavelength of oscillations become longer and longer as we go up the surface, with the last oscillation to emerge from the horizon extending all the way to $t=\infty$. }
\label{matfel}       
\end{figure}

Consider the part of the wavemode that escapes to $r\rightarrow\infty$. In 
fig.\ref{matfel} we have drawn, on the $r-\tau$ plane, the lines of constant phase for this part of the wavemode. We have drawn a timelike surface ($r=constant$) on which we `catch' the mode outside the hole; it will be easier to understand the mode on this surface first and then read off its behavior on any spacelike surface near infinity.

Our goal is to see where the distortion is large enough to create particles. On the initial slice near the horizon, we take a fourier mode $e^{iky^-}$. Consider this mode for the range $-GM<y^-<0$. Recall that $y^-$ is negative outside the horizon, and zero on the horizon. Also, for $y^-\lesssim -GM$, there is no significant distortion of null rays, so that we get $y^-\approx X^-$. What we now wish to show is that even though there is a logarithmic distortion of coordinates for smaller $-GM\lesssim y^-<0$, there is no {\it particle production} for most of this range of $y^-$; in fact particle production will be relevant only for the few oscillations of the wavemode near $y^-=0$.

Thus we now look at the range $-GM<y^-<0$, where we assume that the logarithmic maps (\ref{matqtw}) is a  good approximation. Consider the fourier mode $e^{iky^-}$ on the initial slice. Let the wavelength of this mode be much smaller than $GM$
\begin{equation}
\lambda={2\pi\over k}=\epsilon GM, ~~~~~~~\epsilon\ll 1
\end{equation}
Thus the number of oscillations of the wavemode in our range $-GM<y^-<0$
is large
\begin{equation}
\# ~{\rm oscillations}~=~{1\over \epsilon } \gg 1
\end{equation} 
After the mode evolves to the late time slice we have to look at the wavelength in the  $X^-$ coordinate system.  Consider one oscillation of the wavemode, which in the $y^-$ coordinate system extends over the range
\begin{equation}
-\alpha<y^-<-\alpha+\epsilon
\end{equation}
(Here $\alpha>0$.) The wavelength in the $X^-$ frame will be
\begin{equation}
\lambda_1=|\delta X^-|=|\left ({dX^-\over dy^-}\right )\delta y^-|={GM\over \alpha} \epsilon
\label{matqtoneq}
\end{equation}
Now comes an important question: what about the {\it next} oscillation of the wavemode?  On the initial slice this spans the range
\begin{equation}
-\alpha-\epsilon<y^-<-\alpha
\end{equation}
where we have chosen to look at the oscillation that is the neighboring one on the side closer to the horizon. This evolves to have a wavelength
\begin{equation}
\lambda_2={GM\over |y^-|} \epsilon={GM\over \alpha-\epsilon} \epsilon
\label{matqtone}
\end{equation}
How different is (\ref{matqtone}) from (\ref{matqtoneq})? Let us first suppose that we are {\it not} looking at the first few oscillations of the wavemode near the horizon. Then we have $|y^-|>>\epsilon$, and
\begin{equation}
{\lambda_2\over \lambda_1}\approx {\alpha\over \alpha-\epsilon}\approx 1
\end{equation}
This is the important fact: we can take several adjacent oscillations of the wavemode on the initial slice, and find they evolve to almost the same final wavelength. Thus the stretching they suffer can be called an almost {\it uniform} recaling of coordinates. But under a uniform rescaling we do {\it not} create particles, a fact that we can see as follows. Suppose an initial mode $e^{ik y^-}$ evolves to $e^{ik (\mu y^-)}$, with $\mu$ a (positive) constant. To check for    particle creation we would compute $(f,g)$ where $f=e^{iky^-}$ is the positive frequency mode defining the initial vacuum and and $g=(e^{ik\mu y^-})^*$ is the {\it negative} frequency mode for the final vacuum. But
\begin{equation}
(f,g)=-i\int dy^- e^{ik y^-} e^{ik \mu y^-} \rightarrow 0
\end{equation}
since both fourier modes involved in the integral have the same sign of the exponent. (We get a nonzero integral only if we have $e^{iky}$ and $e^{-iky}$ in the integrand.)

So what we seem to be finding is that the part of the wavemode that is not too close to the horizon undergoes deformations due to the logarithmic stretching, but this does not create particles because under this stretching there is no significant mixing of positive and negative frequency modes. The underlying reason for why we failed to create particles is the same as the analysis of scales that we did in  
the toy model with harmonic oscillators. In the latter case there was no particle creation if the change of the potential was too slow compared to the time period of the oscillator. In the present case, {\it there is very little change in the stretching factor  over the period of oscillaton of the wave}, and so we again get no significant particle creation. 

To make the above conclusion more precise we recall the notion of {\it wavepackets}. 

\subsection{Wavepackets}

\begin{figure}[ht]
\includegraphics[scale=.25]{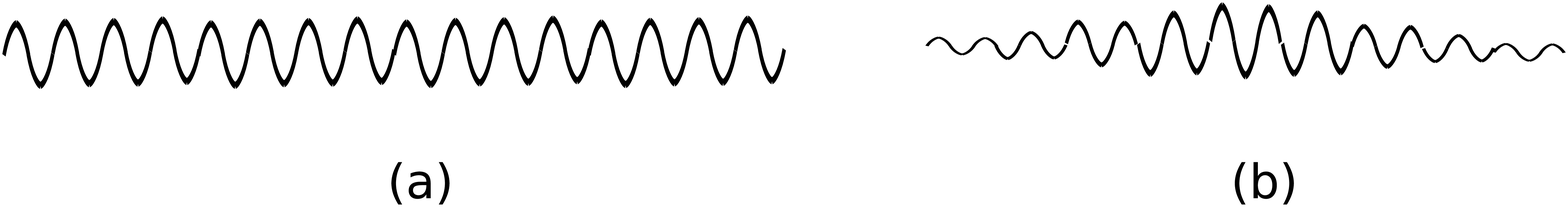}
%
%
\caption{(a) A fourier mode with given wavelength $\lambda={2\pi\over k_0}$ (b) Appropriately superposing fourier modes with wavenumbers near $k_0$ we can make a wavepacket.}
\label{matftw}       
\end{figure}

In fig.\ref{matftw}(a) we depict a wavemode with a definite wavenumber $k_0$. This wavemode has an infinite spatial extent. For physical arguments it is more convenient to have a wavemode that is localized in some region of space. Such a wavemode can be obtained by appropriately superposing wavemodes of different $k$. But we also wish to retain some properties of the mode arising from the fact that the wavenumber was $k_0$. Thus we use only a small band of $k$ around the value
$k_0$
\begin{equation}
k_0-\Delta k < k < k_0+\Delta k, ~~~~~~{\Delta k\over |k_0|} <<1
\end{equation}
This makes a wavetrain that  `sort of' has the wavenumber $k_0$ but which decays after a certain number of oscillations and is thus localized.  Our discussions are mostly qualitative, so we will allow ourselves to use wavetrains that are only a few oscillations long; this means that we will not take  ${\Delta k\over k_0}$ to be very small,  but for our pictorial understanding it will be enough to have $k$ in the rough neighborhood of $k_0$.

\begin{figure}[ht]
\includegraphics[scale=.25]{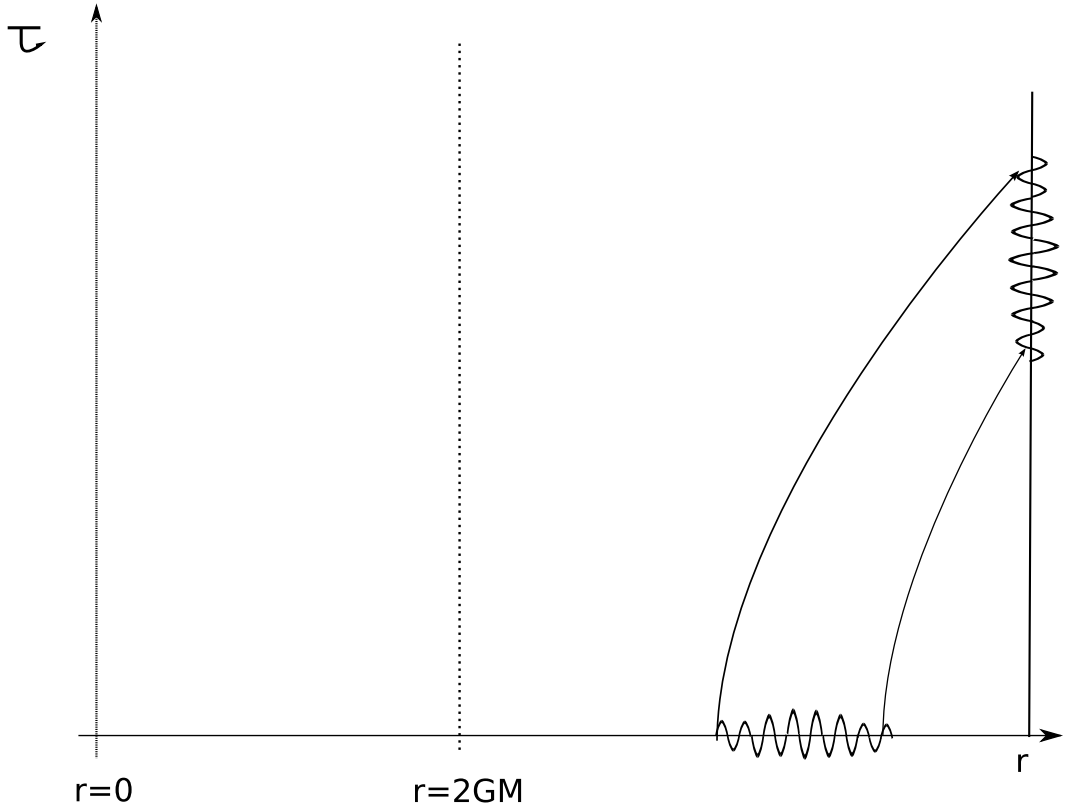}
%
%
\caption{If we look at the oscillations that are not too close to the horizon, then we can make a wavepacket out of them that evolves to a wavepacket at infinity. Suppose we can make a localized wavepacket such that in the  region occupied by the wavepacket the `stretching' of space is approximately uniform. Then there will be no mixing of positive and negative frequencies and therefore no particle production.}
\label{matfthir}       
\end{figure}

Let us now use the above discussion together to make the point that we are after. In fig.\ref{matfthir} we make a wavepacket out of a few oscillations that are  not too close to the horizon. This wavepacket evolves to a wavepacket near spatial infinity without significant distortion, since the oscillations making the wavepacket suffer an almost uniform stretching under the evolution. Thus there is no significant particle production from the part of the wavemode where $|y^-|\gg\epsilon$.

\subsection{Modes straddling the horizon}

So far we have seen what part of the wavemode does {\it not} create particles. The part at $y^-\lesssim -GM$ does not get deformed. The part $-GM\lesssim y^-\ll-\epsilon$ deforms logarithmically but can be broken up into wavepackets each of which suffers `nearly uniform stretching', so again we do not get particle creation.  A similar analysis can be performed for  the domain $y^->0$ which is inside the horizon. We can now turn to the part  of the wavemode that {\it does} create particle pairs. 

\begin{figure}[ht]
\includegraphics[scale=.25]{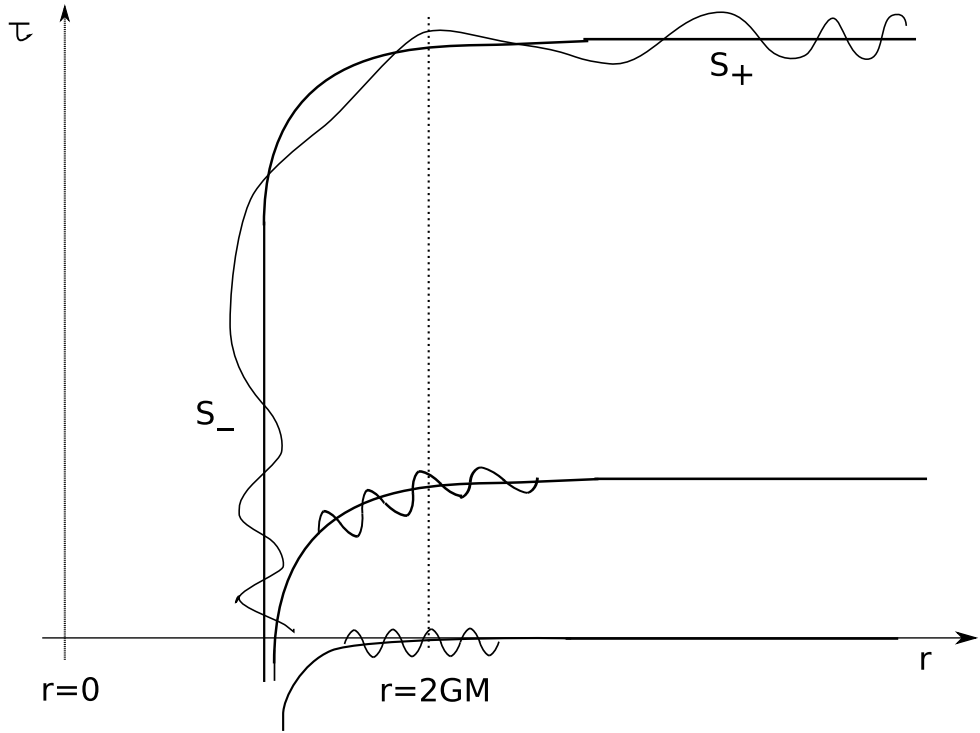}
%
%
\caption{A fourier mode on the initial spacelike surface is evolved  to later spacelike surfaces. In the initial part of the evolution the wavelength increases but there is no significant distortion of the general shape of the mode. At this stage the initial vacuum state is still a vacuum state. Further evolution leads to a distorted waveform, which results in particle creation.}
\label{matffourt}       
\end{figure}

Consider the wavemode on the initial surface and look at the domain of $y^-$ which covers a few oscillations on either side of the horizon $y^-=0$. Thus we have
\begin{equation}
|y^-|\sim \epsilon
\end{equation}
With just a few oscillations in this range, we cannot break this part of the wavemode further into wavepackets. Thus we must evolve it as a whole to the late time surface and see what it becomes. The evolution is described in fig.\ref{matffourt}. On the initial slice we have regularly spaced oscillations. If we look at at surface just a little later, they are still pretty much like regularly spaced oscillations, since there has not been much deformation; thus so far there is no significant particle production. On slices that are much later, we see that the mode has deformed significantly: there are a few oscillations on the part  $S_-$ of the surface that is inside the horizon, then a large gap untill we reach a region on $S_+$, where we find oscillations again.

Note  that on this late time slice the deformation of these oscillations of the wavemode is very {\it nonuniform}. We have a positive frequency mode on the initial surface $e^{iky^-}$, but on the late time surface we will get an admixture of positive frequency modes $e^{iKX^-}$ as well as negative frequency modes $e^{-iKX^-}$. The same happens for the part of the mode on $S_-$. Thus there will be particle creation. 

The most important part of our entire discussion comes now. We know from eq.(\ref{matdone}) that when we create particles by deforming spacetime then the vacuum state changes to a state of the form $e^{-{1\over 2}\sum_{ij}\gamma_{ij} \hat b^\dagger_i \hat b^\dagger_j}|0\rangle$. But in the present case we can break the creation operators $\hat b^\dagger_k$ into two sets: those on $S_+$ which we call $\hat b^\dagger_k$ and those on $S_-$ which we call $\hat c^\dagger_k$. 
When we compute the state on the late time surface it turns out to have the form
\begin{equation}
e^{\sum_k\gamma \hat b^\dagger_k \hat c^\dagger_k}|0\rangle
\end{equation}
We do not derive this result here; the derivation can be found for example in \cite{mathawking,matwald,matparker,matgiddings}. But this is the crucial result for the physics of information,  so we will now spend some time in understanding it. 

\subsection{The nature of the created pairs}

Consider again fig.\ref{matffourt}. On the initial surface the wavemode had a very short wavelength. On later time surfaces the wavelength has been stretched to a longer one, though there is no particle production because the stretching is almost uniform over the oscillations under consideration. The wavelength keeps getting longer as we go to later time slices, till the deformation becomes non-uniform and particles are created. But there is only one length scale in the geometry -- the scale $GM$ -- and one can see easily that when particles are produced the wavelength of the mode has become $\sim GM$. At this point the wavemode has also moved to distances $\gtrsim GM$ from the horizon, and further deformation stops. Thus the wavelength of the produced quanta is $\sim GM$. These are the Hawking radiation quanta, so we see that this radiation has a temperature $\sim \lambda^{-1}\sim {1\over GM}$. The exact temperature is \cite{mathawking}
\begin{equation}
T={1\over 8\pi GM}
\label{matqsone}
\end{equation}
So the wavemode ends its evolution with a wavelength $\sim GM$, but what was its wavelength on the initial slice that we had drawn? On this initial slice there are modes of all possible wavelengths. Consider a wavemode with wavelength {\it shorter} than the one shown in fig.\ref{matffourt}. Then this mode will evolve for a {\it longer} time before it suffers a nonlinear deformation. 

\begin{figure}[ht]
\includegraphics[scale=.25]{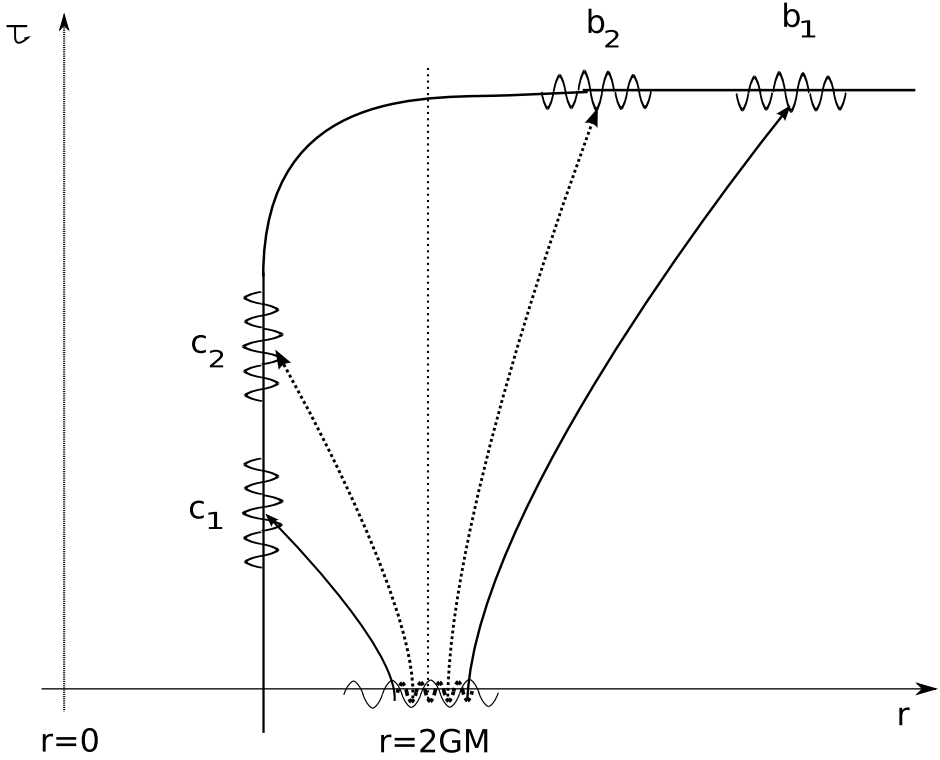}
%
%
\caption{On the initial spacelike slice we have depicted two fourier modes: the longer wavelength mode is drawn with a solid line and the shorter wavelength mode is drawn with a dotted line. The mode with longer wavelength distorts to a nonuniform shape first, and creates an entangled pairs $b_1, c_1$. The mode with shorter wavelength evolves for some more time before suffering the same distortion, and then it creates  entangled pairs $b_2, c_2$.}
\label{matftthree}       
\end{figure}

This situation in depicted in fig.\ref{matftthree}. On the initial slice we have drawn two wavemodes of different wavelengths. The one with the longer wavelength becomes distorted first, and creates the quanta labeled $b_1$ and $c_1$ on the late time slice. The wavemode with shorter wavelength evolves for a longer time before becoming distorted, and  creates the quanta labeled $b_2, c_2$. 

The state of the first pair $b_1, c_1$ is of the form
\begin{equation}
|\psi\rangle_1=Ce^{\gamma \hat b^\dagger_1\hat c^\dagger_1}|0\rangle
\label{matqfone}
\end{equation}
Here $\hat b^\dagger_1$ is an operator that creates a quantum in the localized wavepacket depicted as $b_1$ in fig.\ref{matftthree}, and similarly $\hat c^\dagger_1$ creates the quantum of the wavepacket  labeled $c_1$.  Because we have broken up wavemodes into localized wavepackets, we can define a sort of local vacuum $|0\rangle_{b_1}$ in the region occupied by this mode $b_1$. If we are in this vacuum state then there are no quanta in this region, if we act with $\hat b^\dagger_1$ once then we have one quantum with this wavepacket, if we act with $\hat b^\dagger_1\hat b^\dagger_1$ then we have two quanta of this type, and so on. Doing the same for the modes on $S_-$ we can write the state
(\ref{matqfone}) as
\begin{equation}
|\psi\rangle_1=Ce^{\gamma \hat b^\dagger_1\hat c^\dagger_1}|0\rangle_{b_1}|0\rangle_{c_1}
\label{matqfoneq}
\end{equation}

A similar state is produced by the wavemode which started off with a shorter wavelength on the initial slice. We get particle pairs described by
\begin{equation}
|\psi\rangle_2=Ce^{\gamma \hat b^\dagger_2\hat c^\dagger_2}|0\rangle_{b_2}|0\rangle_{c_2}
\label{matqfoneqp}
\end{equation}

The pairs $b_k, c_k$ for different $k$ lie in regions that do not overlap, so  the overall state on the late time slice is the direct product of the states $|\psi\rangle_k$
\begin{equation}
|\psi\rangle=|\psi\rangle_1\otimes |\psi\rangle_2\otimes |\psi\rangle_3\otimes\dots
\label{matqsixty}
\end{equation}

We have presented a simplified discussion of the created pairs; more technical details can be found in \cite{mathawking,matwald,matparker,matgiddings}. For a more accurate description we should use a large number of oscillations in making each wavepacket (we have used just a few), and then we will have to consider many wavenumbers in each of the intervals on $S_\pm$ over which the wavepackets extend. But the above approximate description has all the essence of what we need to understand the entanglement of quanta.

\subsection{The entangled nature of $|\psi\rangle$}

Consider the state $|\psi\rangle_1$
\begin{eqnarray}
|\psi\rangle_1&=&C\left (|0\rangle_{b_1}\otimes|0\rangle_{c_1} + \gamma\hat b^\dagger_1|0\rangle_{b_1}\otimes \hat c^\dagger_1|0\rangle_{c_1}
+{\gamma^2\over 2}  \hat b^\dagger_1\hat b^\dagger_1|0\rangle_{b_1}\otimes  \hat c^\dagger_1\hat c^\dagger_1|0\rangle_{c_1}+\dots \right )\nonumber\\
&=&C\left ( |0\rangle_{b_1}\otimes|0\rangle_{c_1}+\gamma |1\rangle_{b_1}\otimes|1\rangle_{c_1}+\gamma^2 |2\rangle_{b_1}\otimes|2\rangle_{c_1}+\dots \right )
\label{matqfeight}
\end{eqnarray}
where $|n\rangle_{b_1}$ means that we have $n$ quanta of type $b_1$ in the state etc. 

The important feature of this state is that the $b_1$ and $c_1$ excitations are `entangled'. To understand this in more detail, let us take a simple example of an entangled state.

\subsection{Entanglement and the idea of `mixed  states'}

Consider two electrons, kept at two different locations, and let each of them have a `spin up' state and a `spin down' state. Then this system can have `factored states' of the form 
\begin{equation}
|\psi\rangle=|\psi\rangle_1\otimes |\psi\rangle_2
\label{matqffour}
\end{equation}
Examples are
\begin{eqnarray}
|\psi\rangle&=&|\uparrow\rangle_1\otimes|\downarrow\rangle_2 \nonumber\\
|\psi\rangle&=&{1\over \sqrt{2}}(|\uparrow\rangle_1+|\downarrow\rangle_1)\otimes {1\over \sqrt{2}}(\uparrow\rangle_2+|\downarrow\rangle_2 )
\end{eqnarray}
etc. But we can also have `entangled states which cannot be written as a product of the type (\ref{matqffour}), for example
\begin{equation}
|\psi\rangle={1\over \sqrt{2}}\left ( |\uparrow\rangle_1\otimes|\downarrow\rangle_2+|\downarrow\rangle_1\otimes|\uparrow\rangle_2
\right )
\label{matqffive}
\end{equation}
Suppose we ask: what is the state of electron $1$? For states of type
(\ref{matqffour}) we can answer this question: we ignore the state of electron $2$ and just give the answer $|\psi\rangle_1$. But for states
of type (\ref{matqffive}) we cannot do this, and only the state of the entire system makes sense. Suppose we nevertheless want to ignore electron $2$ in some way. Then we can make a `density matrix'
\begin{equation}
\rho=|\psi\rangle\langle\psi|
\label{matqfseven}
\end{equation}
For the two electron system we get
\begin{eqnarray}
\rho&=&~~{1\over 2}~~ |\uparrow\rangle_1\otimes |\downarrow\rangle_2~~~~
{}_1\langle \uparrow|\otimes {}_2\langle \downarrow|\nonumber\\
&&+{1\over 2}~~ |\uparrow\rangle_1\otimes |\downarrow\rangle_2~~~~
{}_1\langle \downarrow|\otimes {}_2\langle \uparrow|\nonumber\\
&&+{1\over 2}~~ |\downarrow\rangle_1\otimes |\uparrow\rangle_2~~~~
{}_1\langle \uparrow|\otimes {}_2\langle \downarrow|\nonumber\\
&&+{1\over 2}~~ |\downarrow\rangle_1\otimes |\uparrow\rangle_2~~~~
{}_1\langle \downarrow|\otimes {}_2\langle \uparrow|
\end{eqnarray}
We can now `trace over' the states of system $2$, which for the above case means that the bra and ket states of system $2$ must be the same in the terms that we keep. Then we get a `reduced density matrix' describing system $1$
\begin{equation}
\rho_1={1\over 2} ~|\uparrow\rangle_1~~{}_1\langle \uparrow|~~+~~{1\over 2} ~|\downarrow\rangle_1~~{}_1\langle \downarrow|
\label{matqfsix}
\end{equation}
In general we get a density matrix of the form $\rho_1=\sum_{m,n}~ C_{mn} ~|m\rangle_1~{}_1\langle n|$. The probability to find system $1$ in state $k$ is given by the coefficient $C_{kk}$. These probabilities must add up to unity, so we have ${\rm tr} \rho=1$. The {\it entropy} that results from ignoring system $2$ is given by
\begin{equation}
S=-{\rm tr} ~ \rho\ln \rho
\end{equation}
For the density matrix (\ref{matqfsix}) we can compute $S$  easily since it is a diagonal density matrix
\begin{equation}
S=-[{1\over 2}\ln {1\over 2} ~+~{1\over 2}\ln {1\over 2}]=\ln 2
\end{equation}
If the state $|\psi\rangle$ in (\ref{matqfseven}) is `factorized' as in 
(\ref{matqffour}) then when we make $\rho_1$ and compute $S$ then we get $S=0$. Roughly speaking, $S$ gives the log of the number of terms in a sum like (\ref{matqffive}). The entropy is thus a measure of how `entangled' the systems $1$ and $2$ are.

\subsection{Entropy of the Hawking radiation}

Let us now return to the black hole. The state (\ref{matqfeight}) is not factorized between the $b_1$ and $c_1$ excitations. The number $\gamma$ is order unity, so the  first few terms in the sum will be of relevance. To explain the significance of the entangled nature of the state we will for convenience replace the state (\ref{matqfeight}) by the simpler state
\begin{equation}
|\psi\rangle_1={1\over \sqrt{2}}\left ( |0\rangle_{b_1}\otimes|0\rangle_{c_1}+ |1\rangle_{b_1}\otimes|1\rangle_{c_1}\right )
\label{matqfnine}
\end{equation}
The quanta of type $b_1$ lie on the part $S_+$ of the  spacelike surface
which is outside the horizon, while the quanta of type $c_1$ lie on the part $S_-$ which is inside the horizon. Due to the entanglement between $b_1$ and $c_1$ quanta, we cannot restrict ourselves to the Hawking radiation quanta $b_1$ and still describe them by a `pure' quantum state. If we wish to ignore the quanta $c_1$ then we have to find the density matrix for the quanta $b_1$. For the state (\ref{matqfnine}) we will get an entanglement entropy $S=\ln 2$. (The state (\ref{matqfeight}) would have given an $S$ of the same order.)

Now we can look at the other pairs of quanta $(b_2, c_2), (b_3, c_3) \dots$. We had seen that each of these sets $(b_k, c_k)$ lives at a location different from the other pairs, so the overall state (\ref{matqsixty}) was a direct product of states for each of these pairs. A little thought shows that the total entanglement entropy $S$ will then be  the sum of the entropies from each pair 
$(b_k, c_k)$. Let us see how many such pairs there will be. The temperature of the Hawking radiation is (\ref{matqsone}), so the energy of the typical emitted quantum is $\sim (GM)^{-1}$. The mass of the hole is $M$, so the number of quanta that will be emitted when the hole has evaporated is 
\begin{equation}
\#~quanta\sim M (GM)\sim {(GM)^2\over G}
\label{matqssix}
\end{equation}
With an entropy of order unity from each set $(b_k, c_k)$ we see that the entropy of the radiation is
\begin{equation}
S_{rad}\sim {(GM)^2\over G}\sim S_{bek}
\label{matqstwo}
\end{equation}
so we see that the radiation has an `entanglement entropy' of the order of the entropy of the black hole. 

\subsection{The problem with the entangled state}

Consider the two-electron state (\ref{matqffive}), and suppose that we want to concentrate on the first electron. We have seen that we cannot write a quantum state for this electron alone. We can make a density matrix $\rho_1$, but this is not a `pure' quantum state. Rather it is a statistical construct that allows us to get probabilities for different states of electron $1$, and one cannot see the usual quantum principles of linear superposition or phase interference by looking at $\rho_1$.

Of course there is no fundamental problem with such an entangled state; all we have to do is realize that it is only the complete 2-electron system that can be described by a quantum state. The situation is a bit different for the black hole case. As long as we are willing to look at both sets of  quanta,  $b_k$ and  $c_k$, then we have an entangled quantum wavefunction. But if the black hole eventually disappears, then we will be left with the quanta of type $b_k$ floating at infinity. We know that they cannot be described by a pure quantum state, and now we cannot write a mixed state either, for there is nothing for them to mix with! Thus the only way we can describe the $b_k$ quanta is by the reduced density matrix $\rho_b$ describing the $b_k$, and this description is inherently statistical, rather than a usual quantum mechanical one. This is what led Hawking to postulate that quantum mechanics in the presence of gravity is not a consistent theory by itself; he suggested that general configurations can only be described by density matrices, and we must make a quantum theory based on such a description.

Attempts to modify quantum theory in this way have not made much progress.  Others have argued that the black hole does not completely evaporate away, but instead stabilizes after reaching planck size because of quantum gravity effects. In this case the quanta $c_k$ are never removed from the system, and we have a pure state overall. But one is then forced to accept that there can be an infinite number of possible states of such a planck sized remnant (since the remnant can result from an arbitrarily large black hole). Allowing the theory to have infinitely many states within a bounded spatial region and within a bounded energy range is unnatural, and creates many problems for the theory. It would therefore seem best if somehow we could get the black hole to disappear and yet have the quanta $b_k$  be left in a pure state. Let us now discuss what would be needed for this to be possible. 

\section{Common misconceptions about information loss}

We will find it helpful to start by considering several common misconceptions about how information can come out of the black hole.

\subsection{Is the emitted radiation exactly thermal?}

A common argument about Hawking radiation is the following.  The above discussed computations give `thermal radiation',   but there could be corrections (from the gravitational backreaction of the created pairs, for example) which generate  small deviations from `thermality',  and these deviations can encode the information that should escape from the hole.

\begin{figure}[ht]
\includegraphics[scale=.25]{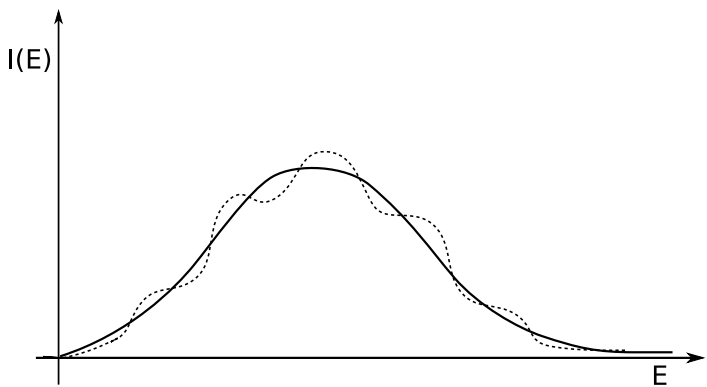}
%
%
\caption{The planck distribution; small deviations from this distributions are indicated by the dotted curve.}
\label{matfeightt}       
\end{figure}

The problem here is the word `thermal'. What is thermal radiation? One might think that `thermal' means the spectrum of radiation should be plankian; this spectrum is depicted by the solid curve in fig.\ref{matfeightt}. Small deviations from this spectrum are shown by the dotted curve in fig.\ref{matfeightt}. Can such a change in spectrum  bring out information from  the black hole? We will now see that the shape of the spectrum itself does not have much to do with whether the information comes out.

For one thing, the spectrum of the semiclassical radiation from the black hole is  not of the planck shape; the spectrum is modified by greybody factors. This is a general feature of radiation from any warm body -- there is a modification to the spectrum if the emitted wavelength is comparable to the size of the body. For black holes, this wavelength is $\sim GM$, which is the same order as the black hole size $r\sim 2GM$. Thus the spectrum is not planckian anyway.

A more correct definition of 	`thermal' radiation is that if the body has an absorption cross section $\sigma(k)$ for quanta of a certain wavenumber, then the emission rate for the same wavenumber is
\begin{equation}
\Gamma=\sigma(k) {d^3k\over (2\pi)^3}{1\over e^{\omega\over T} -1}
\end{equation}
The semiclassical radiation from the hole is `thermal' in this sense. But the essential problem that we have is {\it not} created by this `thermality', but by the entangled nature of the state. Whether we have the entangled state (\ref{matqfeight}) (which can be shown to be `thermal' in the above sense) or the entangled state (\ref{matqfnine}), which is very different from `thermal', we face the {\it same} problem. There is order unity entropy of entanglement from the state created by each pair of operators $(\hat b^\dagger_k, \hat c^\dagger_k)$, and so there is an entanglement entropy (\ref{matqstwo}) for the radiation which is order $S_{bek}$. It is this entanglement that will eventually lead to information loss. By contrast, if a piece of coal burns away completely to radiation, then this radiation is in a pure state, even though it looks much more `thermal' than a state which has the form (\ref{matqfnine}) for each of the $(\hat b^\dagger_k, \hat c^\dagger_k)$.

Thus `thermality' is not really the issue; the issue is the entangled nature of the state created in the process of black hole evaporation.

\subsection{Can small quantum gravity effects encode information in the radiation?}

Consider the derivation of Hawking radiation discussed in the above sections. We have used a classical metric and a quantum field $\phi$ on this `curved space', but gravity itself has not been treated as quantized; this is called the {\it semiclassical approximation}.
Thus the semiclassical computation of radiation does not use the physics of quantum gravity anywhere. Since spacetime curvature was low in the regions where the wavemodes deformed and created particles, this would seem to be a good approximation. But one can still wonder if the {\it small} corrections that would arise from quantum gravity effects 
could change the state of the radiation to a pure state. There are two aspects to this question:
 
\bigskip

(a) The first point to note is that a {\it small} change in the state of the quantum field will {\it not} succeed in making the state of the $b$ quanta a pure state. Focusing again on a given set $(b_1,  c_1)$ we see that their state is a mixed one like (\ref{matqfeight}). To get no entanglement of the $b_1$ quanta with the $c_1$ quanta we would need a state like
\begin{equation}
|\psi\rangle_1=\left ( C_0|0\rangle_{b_1}+C_1|1\rangle_{b_1}+\dots\right )\otimes \left (    D_0|0\rangle_{c_1}+D_1|1\rangle_{c_1}+\dots\right )
\label{matqsthree}
\end{equation}
But the state (\ref{matqsthree}) is {\it not} a small perturbation on a state like (\ref{matqfeight}). The two states are completely different, so we  need an order {\it unity} change in the state of each set $(b_k, c_k)$ before the state can become pure.  Thus if quantum gravity is to help us, {\it then it must completely change the evolution of the wavemodes that we have been drawing in the above sections}.

\bigskip

\begin{figure}[ht]
\includegraphics[scale=.20]{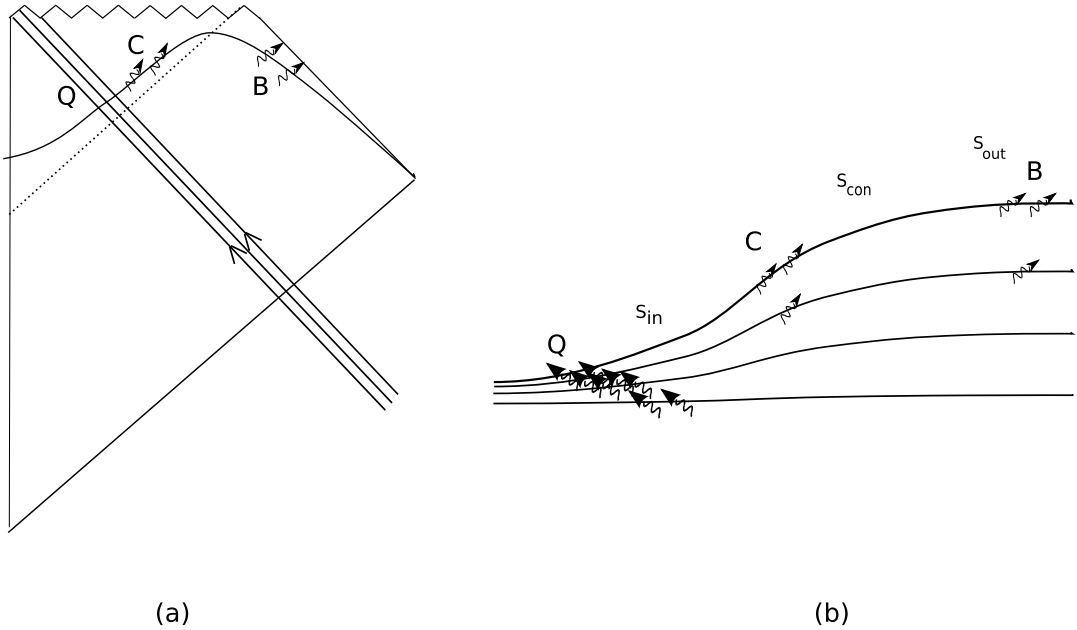}
%
%
\caption{(a) The infalling matter $Q$ and the entangled pairs $C,B$ shown on the spacelike slices in the Penrose diagram. (b) $Q,C,B$ sit at different locations on the spacelike slices. To catch all three of these on the slices while staying in a low curvature region we have evolved the small $r$ side less and the large $r$ side more, something that we are certainly allowed to do in classical gravity.}
\label{matfsevent}       
\end{figure}

(b) The second point is that even if we had a state like (\ref{matqsthree}), and thus the radiation quanta $b_k$ formed a pure state by themselves, it would not solve the information problem. Consider the Penrose diagram in fig.\ref{matfsevent}(a). There are not two but {\it three} kinds of matter involved in the problem. There is the matter that fell in to make the hole, marked $Q$. Then there are the Hawking radiation quanta $b_k$ (we have labeled them $B$) and their entangled partners, the $c_k$ (labeled $C$ in the figure).

The problem is that not only do the quanta $B$ have to form a pure state, they have to carry the information of the matter $Q$. This is because in quantum mechanics the evolution of states is one-to-one and onto, and so different states of the initial matter $Q$ have to give different states of the final radiation $B$. In fig.\ref{matfsevent}(b) we have drawn the slices as shown in fig.\ref{matffour}, with $Q, B, C$ indicated. We see that the quanta $Q$ reach small $r$ first, and exist on each slice. The way we have drawn our slices keeps $Q$ always in a region of low curvature; to achieve this we have evolved the small $r$ region very little as we move from slice to slice. As the evolution proceeds the $b_k$ and $c_k$ quanta start appearing out of the vacuum modes.  But these vacuum modes were localized in the region between the $b$ and $c$ quanta, far away from where $Q$ sits on the slice. So  {\it how can the matter $Q$ transfer its information to the $b_k$?} This is the essence of the information problem. 

Note that all the evolution depicted in fig.\ref{matfsevent}(b) has been in a low curvature region, with slices that are smooth and carrying matter that is always low density. Thus it would appear that the situation is like the low curvature physics encountered in the solar system, and no unexpected quantum effects can occur. The only unusual thing is that through the course of the the evolution the slices {\it stretch} by a large amount, as discussed in section \ref{matsecslice}. In conventional  relativity the total stretching from initial to final slice does {\it not} matter; quantum gravity effects will not come in  as long as the {\it rate} of change is small. This fact may not be true in string theory; for a discussion see \cite{matstretch}. 

\subsection{What is the difference between Hawking radiation and radiation from a burning  piece of coal?}

Suppose a piece of coal burns away completely, leaving behind only the radiation it emitted. This time we know that subtle correlations in the emitted quanta encode the entire information about the state of the coal. But because these correlations are subtle, we cannot see them easily. How does this radiation differ from the Hawking radiation emitted by  the black hole?

Consider the first photon emitted by the coal. This photon {\it can} be in a mixed state with the matter left behind in the coal. Let us assume that an atom emits this photon, and that after the emission the spin of the atom and the spin of the photon are correlated in an entangled wavefunction as follows
\begin{equation}
|\psi\rangle_1={1\over \sqrt{2}}\left ( |\uparrow\rangle_a\otimes |\downarrow\rangle_{p}+|\downarrow\rangle_a\otimes |\uparrow\rangle_p\right )
\end{equation}
where $ |\uparrow\rangle_a$ stands for the spin up state of the atom,  $|\downarrow\rangle_{p}$ stands for the spin down state of the emitted photon, etc.. Thus far, the situation looks just like the case of entangled $b,c$ in the  black hole. But the crucial difference is that when later photons are emitted from the coal, they can bounce off the atom left behind in the coal, and thus the spins of these later photons can carry the information left behind in this atom. If this atom  drifts out itself (as a piece of ash) then it can also carry the information of its spin. Thus at the end the quanta collecting at infinity are entangled only with themselves, and form a pure state carrying all the information in the initial piece of coal.  

Contrast this with the state of the radiated quanta $b_k$ in the black hole case, shown in fig.\ref{matftthree}. The quanta of type $b_1$ are correlated with the quanta of type $c_1$, which are located at a certain region on the part  $S_-$ of the spacelike slice.  {\it But this place where $c_1$ is located is not involved any further in the process of radiation from the black hole.}  
For example,  consider a later pair , say $(b_{10}, c_{10})$, and look at the region where this mode is suffering its  nonuniform deformation.  This region is {\it not} causally connected to the location where the earlier quanta $c_1$ is located, so $c_1$ cannot have any influence on the later emitted quantum $b_{10}$. In the case of the coal the atom left behind after the first emission {\it was} in causal contact with later quanta leaving the coal. The black hole is different because each pair $(b_k,c_k)$ is created at a point on a spacelike surface, and then this surface {\it stretches} so that the $b_k, c_k$ quanta are moved away in different directions.  New quanta are again created in the middle (i.e. at the horizon), these  are again moved away by stretching, and so on. Thus all the created quanta $b_k, c_k$ are located along different points of a very long spatial slice, with no overlap in their locations. 

Since the quanta are prevented from influencing each other by being spread out along this very long spatial slice, we should ask the basic question: how did we get this very long spatial slice when the black hole only had a given size $\sim GM$? Recall from fig.\ref{matfthree} that the spacelike slice inside the horizon was of the form $r=constant$, and it could be made arbitrarily long while remaining in the region $r<2GM$. This possibility is unique to the black hole geometry, since it needs the light cones to `turn over' and make the $r=constant$ direction spacelike. This  does not happen for the coal, and so later quanta can (and do) carry the information left in entangled pairs from earlier quanta.

\section{The Hawking `theorem'}\label{theorem}

There is one more common misconception about Hawking's computation of radiation which is very important to address. Look at the evolving mode drawn in fig.\ref{matffourt}. On the late time surface this mode was deformed, but if we follow the mode to the far {\it past} then it is just a simple fourier mode  with no particles in that mode; i.e. $\hat a_k|0\rangle=0$. The further back we look, the smaller the wavelength. In fact if we follow the mode to times before the black hole formed then we find that its wavelength was much shorter than planck length; such modes are called `transplanckian'. But perhaps we do not really know how to do quantum field theory  when transplanckian wavelengths are concerned. In normal physics we take a field, break it into fourier modes, make operators $\hat a_k, \hat a^\dagger_k$ and define a vacuum annihilated by the $\hat a_k$. Maybe all this is incorrect when describing transplankian modes, and quantum gravity must be brought in in some essential way?

If this argument were correct, then we have no information paradox, since Hawking's semiclasical computation would be invalid. In this section we argue that we do {\it not} need to know the physics of transplankian modes to make Hawking's claim; we can formulate his argument using only physics at  scales that we understand. More precisely, we will formulate his argument in the form of the following `theorem':

\bigskip

Suppose we are given that

\bigskip

(a) The effects of quantum gravity are confined to within a fixed length like planck  $l_p$ or string length $l_s$.

\bigskip

(b) The vacuum is unique.

\bigskip

Then when a black hole forms and evaporates, we {\it will} have information loss.

\bigskip

The meaning of these conditions will become clearer as we go through the argument.

\subsection{The local vacuum}

Let us first chose a length scale where we believe that we {\it do} understand quantum field theory and its vacuum structure. This could be $\lambda\sim 1~ {fermi}$, since experiments on nuclear scales agree well with computations of feynman graphs in field theory. Or we could take $\lambda\sim 1~{Angstrom}$, since we understand atomic physics well, including the effects of vacuum fluctuations in effects like the Lamb shift. It does not matter what scale we choose; we will just keep it fixed henceforth as a scale $\lambda\sim \lambda_{known}$. The black hole itself will be taken as big, so we have 
\begin{equation}
l_p\ll \lambda_{known}\ll GM
\end{equation}
A given fourier modes starts off with very small wavelength $\lambda\ll l_p$, evolves to longer wavelengths $\lambda\sim \lambda_{known}$, and then continues to evolve to $\lambda\sim GM$, where its distortion becomes nonuniform and particle pairs are created. The important point is that since $\lambda_{known}\ll GM$, no  particle pairs  have been created when  $\lambda \sim \lambda_{known}$. Thus we will look at the physics at this intermediate scale $\lambda_{known}$ which is much larger than planck length and where we the wavemode is still in the vacuum state.

\subsection{The consequences of conditions (a) and (b)}

 Look at the region circled in fig.\ref{matfeight}(a). If we assume condition (a) of our `theorem',   then since the circled region is far from the singularity we have `normal' physics' in this region, with no quantum gravity effects. That is, the metric is that of  empty, almost flat,  spacetime. Now focus on a mode which in this region has $\lambda\sim \lambda_{known}$. By condition (b) of the `theorem' the vacuum is unique, which means that there is only `one kind of empty space' possible in the theory; this empty space must therefore be described by the usual quantum vacuum that we use in field theory.  
Since there is nothing strange about the state of the spacetime region under consideration, the fourier  mode that we are studying (with $\lambda\sim \lambda_{known}$) will have to behave the way we expect a mode to behave in usual field theory. 

 Since we have `normal physics' for this mode, the possible states of this mode are the vacuum $|0\rangle$, 1-particle $|1\rangle$, 2-particle $|2\rangle$, etc. There are now two possibilities:
 
 (i) First assume that the state of the field mode is the vacuum $|0\rangle$. Then the state will evolve in the way shown in fig.\ref{matffourt}. So the mode will become distorted and create entangled particle pairs described by a state like
(\ref{matqfeight}), and we would have the information problem created by such an entangled state.

(ii) What happens if we assume that the state of the mode $\lambda$ was {\it not} the vacuum state in this circled region? It is in principle possible that we  get such an excited state for the mode because as we have argued above,  we do  not really know the evolution of the mode at the time when it was transplanckian.  So suppose the mode is in a 1-particle state $|1\rangle$ when it reaches $\lambda\sim \lambda_{known}$. Then because for this mode we have `normal physics', we will have the energy density expected  from quanta of $\lambda\sim\lambda_{known}\sim 1~{fermi}$ in the circled region of fig.\ref{matfeight}. So there would be matter of {\it nuclear} density filling this region. This would {\it not} agree with this region being low curvature `empty space',  as required by postulate (a). More generally, the state of the wavemode $\lambda\sim \lambda_{known}$ can be
\begin{equation}
|\psi\rangle=C_1|0\rangle+C_1|1\rangle+C_2|\rangle+\dots
\label{matqsfive}
\end{equation}
If $C_i, i>0$ are not small, then we get the nuclear density matter distribution around the horizon. (It does not help to ask that the $C_i$ be small but nonzero, since then the evolved state will be close to (\ref{matqfeight}), and we have already seen that we need an order {\it unity} change in this state to remove the entanglement.)

\subsection{The consequence of a non-unique vacuum}

It may appear that there is one way that we can have the classical geometry of the hole depicted in the circled region of fig.\ref{matfeight}(a), and yet avoid information loss. This way would be to drop condition (b) from our set of natural physics assumptions. Let us see what dropping this condition  would imply.

  Consider again the state of the quantum field in the circled region of fig.\ref{matfeight}(a). Suppose that the state here is {\it not} the usual vacuum, and yet it has {\it no} energy density. This sounds strange, and indeed there are no such states in usual field theory. But it could be that the transplankian modes, which we do not understand, have some complicated states which are not the usual vacuum, and yet have no extra energy over that of the vacuum. Then the evolution of   modes with $\lambda\sim \lambda_{known}$ {\it can} be different from the  normally expected evolution  because of interaction with these `hidden' transplanckian excitations. The allowed states for modes $\lambda\sim \lambda_{known}$  may not be of the form 
(\ref{matqsfive}), and the evolution of these modes may not be the usual free wave evolution depicted pictorially in fig.\ref{matffourt}. 

But if such a situation were permitted in our full quantum gravity theory, then we would have to say that the vacuum of the theory is non-unique. There would be an arbitrarily large number of states possible in a given region with energy arbitrarily close to the vacuum. For each such state we would find a  totally different evolution for modes with $\lambda\sim \lambda_{known}$. In this situation the theory loses all predictive power. In the lab  we would not know which of these `vacuum' states we have, so we would not know how modes with $\lambda\sim \lambda_{known}$ would behave. 
We could never do the physics at any length scale, because modes with shorter length scales could be `corrupting the vacuum' and modifying evolution, without being detectable since they contribute no net energy. Thus we normally assume condition (b) of our theorem: that the vacuum {\it is} unique. (For an example of a theory with a non-unique vacuum created by nonlocal identifications, see \cite{matmc}.)

\subsection{Summary of the information paradox}

Thus we see that if we assume the two very reasonable sounding assumptions (a), (b) of the Hawking `theorem', then we are forced into a situation where the outgoing radiation will not be a pure state carrying the information of the black hole. To evade the information paradox we will therefore need some radical change in our basic understanding of quantum mechanics and gravity. Let us first summarize the main ideas that have led to the information paradox.

The central point is that vacuum modes evolve over smooth spacetime in the manner sketched in fig.\ref{matffourt}, and thus create entangled particle pairs. Entangled states are not a problem by themselves.  The problem arises because gravity is an attractive force with a negative potential energy, and this makes the quanta $c_k$ inside the horizon have a net negative energy. Thus the matter $Q$ and the quanta $C$ in fig.\ref{matfsevent} can have a net mass zero. Then all the energy will go the the $b_k$ quanta and there is no net mass left in the hole. If we assume that there cannot be an infinite number of light `remnants' in our theory then we are forced to assume that the black hole disappears. Now the radiation quanta $b_k$ are `entangled with nothing', and we cannot describe them by any wavefunction.

To save this situation we need some way to {\it change significantly} the evolution depicted in fig.\ref{matffourt}. In fact what  we need is not only that the matter labeled $B$ in fig.\ref{matfsevent} be in a pure state (so that it should not be entangled with $C$), but that it should reflect all the information in the matter $Q$. In the derivation of the Hawking `theorem' we saw that we could restrict attention to wavemodes with $\lambda\gtrsim \lambda_{known}$, where the physics of evolution is well understood. The evolution of these modes, depicted in fig.\ref{matffourt}, would seem to be governed by physics that we know very well -- the physics of quantum fields on gently curved space. {\it Yet, to save quantum theory we need that this evolution be changed by order unity, leading to a completely different state than the entangled pair state that we got!}  A small change in the evolution, leading to a small change in the final state, will not help.  

We will see in the next section that in string theory it is condition (a) that fails; quantum gravity effects can change the entire interior of the hole and resolve the information paradox.

\section{Black holes in string theory: fuzzballs}

String theory provides a consistent theory of perturbative quantum gravity, so we can hope that the theory might also be able to avoid contradictions when it comes to nonperturbative things like black holes. The theory has no free parameters, and no fields can be added or removed from the theory. To make the the black hole we must use the objects present in the theory. Let us compactify the 10-d spacetime of string theory as follows
\begin{equation}
M_{9, 1}\rightarrow M_{4,1}\times T^4\times S^1
\label{matqsseven}
\end{equation}
We can wrap a string around the $S^1$; this will look like a point mass from the viewpoint of the noncompact directions. We can take a large number $n_1$ of these strings  and ask what metric they produce. The important thing is that we take a {\it bound}   state of the strings, otherwise we will make `many small black holes' rather than the one massive hole that we are seeking. The bound state of these strings is easy to picture: the string just wraps $n_1$ times around $S^1$ before closing. There is just one such state of the string, since the string is an `elastic band' and settles down to its shortest length for the given winding. Thus the microscopic count of states would suggest an entropy $S_{micro}=\ln 1=0$. What about the `black hole' that it creates? The string carries `winding charge' and radiates a corresponding 2-form gauge field $B_{\mu\nu}$. When we make the metric with the mass and charge of the string we find that the horizon coincides with the singularity, and so the horizon area is zero. Thus the Bekenstein entropy $S_{bek}={A/4}=0$, and so we get $S_{bek}=S_{micro}$.

Alternatively we can  take the massless gravitons of the theory and allow them to circle around the $S^1$; this would also look like a mass point from the viewpoint of the noncompact directions, but now the mass point will carry `momentum charge' due the momentum carried by the gravitons.  To get a `bound state' of these gravitons we would have to put all the momentum into one energetic graviton, so the microscopic entropy would be again $S_{micro}=\ln 1=0$. The metric produced by this graviton carrying energy and `momentum charge' again ends up with no horizon area, and we get $S_{bek}=0=S_{micro}$.

To get something more interesting let us {\it combine} the winding and momentum charges. To make a bound state of winding and momentum we simply let the momentum be carried as traveling waves on the string. But now we see that there are many states for a given winding $n_1$ and a given momentum $n_p$: we can put all the energy in the lowest harmonic, or some in the first and some in the second harmionic, or take any other distribution of the energy into harmonics. The number of such states turns out to give an entropy \cite{matsen}
\begin{eqnarray}
T^4: ~~~S&=&2\sqrt{2}\pi\sqrt{n_1n_p}\nonumber\\
K3:~~~S&=&4\pi\sqrt{n_1n_p}
\label{matqsnine}
\end{eqnarray}
where we have also included the answer for a case where the $T^4$ in (\ref{matqsseven}) has been replaced with another 4-d manifold called $K3$.

We can compute  the geometry produced by a point source carrying the energy and gauge fields produced by the string winding and    momentum. In this computation we should note that the string action contains $R^2$ corrections to the leading Einstein action $R$. This modifies the expression for the Bekenstein entropy (to the `Bekenstein-Wald entropy' \cite{matwi}). With these needed corrections this entropy has been computed for the case of K3 compactification, and one finds \cite{matdabholkar}
\begin{equation}
S_{bek}=4\pi\sqrt{n_1n_p}=S_{micro}
\end{equation}
so the microscopic count exactly reproduces the entropy from the geometry of the horizon. 

\begin{figure}[ht]
\includegraphics[scale=.25]{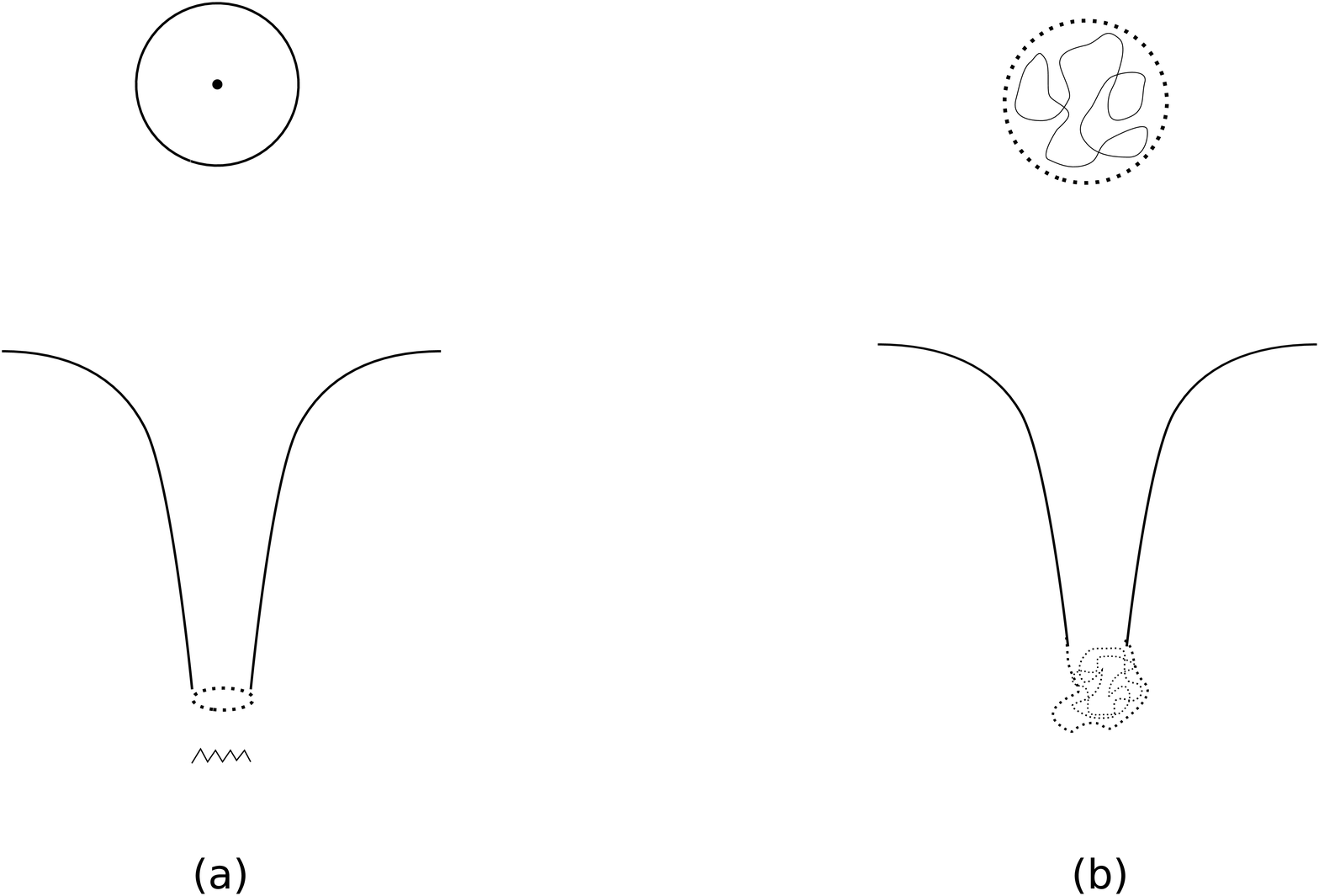}
%
%
\caption{(a) If the string winding and momentum excitations could sit at a point, then we would get the usual black hole; in the lower diagram the geometry is shown with flat space at infinity, then a `throat', ending in a horizon with a singularity inside. (b) The string cannot carry the momentum without transverse vibrations, and thus spreads over a horizon sized transverse area. The geometry depicted in the lower diagram has no horizon; instead the throat ends in a `fuzzball'.}
\label{matftfive}       
\end{figure}

We can make more complicated holes, by adding $n_5$ 5-branes wrapped on $T^4\times S^1$ (or on $K3\times S^1$). This time the horizon area is large enough that we do not need the $R^2$ corrections to the action, and one finds an exact agreement again with the microscopic count of states \cite{matsv}
\begin{equation}
S_{bek}={A\over 4}=2\pi\sqrt{n_1n_pn_5}=S_{micro}
\label{matqseight}
\end{equation}
So we seem to understand something about black hole entropy, but what about the information problem? To understand what can change in Hawking's derivation of information loss, we need to understand what is going on inside the black hole. Let us return to the 2-charge hole made with string winding and momentum. The crucial point is that the elementary string of string theory has no {\it longitudinal waves}; it admits only transverse oscillations. Thus when carrying the momentum as traveling waves it spreads over some transverse region, instead of just sitting at a point in the noncompact space. Instead of the spherically symmetric hole with a central singularity at $r=0$ we get a `fuzzball', with different states of the string creating different fuzzballs. Interestingly, the boundary of the typical fuzzball has an area that satisfies
\begin{equation}
{A\over G}\sim \sqrt{n_1n_p}\sim S_{micro}
\end{equation}
So we see that the region occupied by the vibrating string is of order the entire horizon interior; in fact a horizon never forms \cite{matstring}. We depict this situation in fig.\ref{matftfive}.
Now there is no information problem: any matter falling onto the fuzzball gets absorbed by the fuzz, and is eventually re-radiated with all its information, which is just how any other body would behave. The crucial point is that we do not have a horizon whose vicinity is `empty space'. The matter making the hole, instead of sitting at $r=0$, spreads all the way to the horizon.  So  it can send its information out with the radiation, just like  a piece of coal would do.

Such constructions have been carried out for a large class of microstates \cite{matmany}, including states of the 3-charge hole carrying winding, momentum and 5-brane charges, and more complicated holes with four kinds of charges. Some states of non-extremal holes have been made as well \cite{matross}. Radiation from these non-extremal gravity states has been computed \cite{matmyers}, and found to agree exactly with the radiation expected from the corresponding state of strings and branes \cite{matcm}.

We can still ask: why does all this work? What feature of string theory led to this large change in the picture of the hole, and allowed the interior of the horizon to depart from the naive classical expectation? The answer would seem to be `fractionation', a phenomenon peculiar to string theory which is a theory of {\it extended} objects. Consider spacetime with a compact circle of  length $L$. Suppose we want to make an excitation of this system, while adding no net charge. What is the lowest energy $\Delta E$ that we will need? We can take one graviton in the lowest allowed harmonic running clockwise on the $S^1$,  and one running; this would give an energy $\Delta E={4\pi\over L}$. Now suppose on this circle we already had a wrapped string with winding $n_1$. Now we can excite a clockwise momentum mode of energy ${2\pi\over n_1L}$ on the string, and with a similar contribution from the anticlockwise mode we get $\Delta E={4\pi\over n_1 L}$.  If $n_1\gg 1$ then this $\Delta E$ is much smaller than the energy gap in the absence of the strings.  We say that in the presence of the strings the momentum comes in {\it fractional} units, which are ${1\over n_1}$th of a full unit \cite{matdmpre}. 

This looks like a simple physical effect, so what can it have to do with black holes? In string theory we have duality, which allows us to map different objects in the theory to each other. Thus we can map the $n_1$ times wound string to a bound state of $n_1$ 5-branes. At the same time the momentum mode would map to a string winding along the $S^1$. Now the `fractional momentum mode' becomes a `fractional string'. But what is a fractional string? The original string had a tension of string scale, which is order planck scale. But the fractional string has a tension which is ${1\over n_1}$th of this value, and so for $n_1$ large it will be a very low tension object \cite{matmaldasuss}.

One can extend such constructions further, to bound states of many kinds of branes. Let us take the black hole described in (\ref{matqseight}). One finds that there exist very low tension `floppy fractional objects'  that  stretch over  distances of order \cite{matemission}
\begin{equation}
D\sim \left [ {(n_1n_pn_5)^{1\over 2} g^2 \alpha'^4\over VL}\right ] ^{1\over 3}
\label{matqqq}
\end{equation}
where $V$ is the volume of $T^4$, $L$ is the length of $S^1$ and $g,\alpha'$ are the string coupling and tension. But this turns out to be  just the order of the horizon radius of the black hole with these charges! This argument tells us that fractionation can generate quantum effects over horizon scales. We can then return to simpler holes like the 2-charge hole (\ref{matqsnine}), where we can construct the internal state of the hole, and see that we indeed get a `fuzzball' instead of the traditional hole.

This solves the information paradox, but raises many natural questions about the behavior of black holes. While the dynamics of fuzzballs is in its infancy, we can make some simple observations and conjectures relevant to such questions.

\bigskip

{\it If a shell of dust is collapsing, will it suddenly change its dynamics when it reaches horizon size?} 

No, the fuzzball proposal does not require that. The essential point is that there are {\it two} timescales in the black hole problem. One is the `crossing time scale' of order $\sim GM$, over which the collapse occurs. The other is the much longer Hawking evaporation timescale, $t_{evap}\sim GM( {M\over m_{plank}})^2$. The collapsing matter was in a low entropy state, and will take some time to come to statistical equilibrium and  reach a generic state (which we expect to be a fuzzball-type state). It is known that the entropy of radiation from the hole $S_{rad}$  is somewhat larger than $S_{bek}$, since the radiation free-streams out of the hole rather than leave in a `quasi-static' way \cite{matpreskill}. Thus the matter can collapse as classically expected on the crossing time scale, and even use some fraction of $t_{evap}$ to stabilize to the fuzzball configuration; we can still carry the information out in the remaining radiation. 

\bigskip

{\it After the black hole has stabilized to the fuzzball configuration, will an infalling  body feel a very different environment from that of the usual black hole geometry?} 

Not necessarily, since the `fuzz' is a very low density `web', at least in the simple 2-charge examples that we can explicitly study \cite{matweb}. 
If a body is heavy (compared to the energy of a Hawking radiation quantum) {\it and} we follow it only over the short `crossing' timescale $\sim GM$, then we may not see a dynamics that departs significantly from the classical one. But over the long Hawking evaporation timescale the information in the heavy body should get incorporated in the fuzz and eventually get radiated away.

\bigskip

{\it After the black hole has stabilized to the fuzzball configuration, will the evolution of Hawking radiation quanta be very different from that expected in the classical geometry?} 

Yes, and that {\it should} happen. If we do not modify the evolution of  $\lambda\sim GM$ fourier modes in the vicinity of the horizon, we will have information loss, as argued in the above sections. The fuzzball structure of the hole ensures that the information of the hole  reaches out to the boundary of the hole and so the mode evolution of fig.\ref{matffourt} is altered, not slightly, but rather by order unity effects. This is what is needed to prevent information loss.

\section{Conclusion}

So what is the information paradox? We would like ordinary quantum theory to be valid, even when black holes form and evaporate. But with the traditional picture of the black hole, the explicit computation of Hawking radiation generates entangled pairs, and the state of the outgoing quanta is not a pure quantum state when the black hole disappears. Further,  the state of these outgoing quanta $b_k$ has no relation to the matter that made the hole; they just made a specific entangled state with their partners $c_k$. {\it To resolve the paradox we have to find some way to change the evolution of vacuum modes depicted in fig.\ref{matffourt}, so that the $b_k$ form a pure state containing the information of the initial matter.} Small changes in the evolution will not help; it has to be an order unity change since we want a completely different outcome. But if we make some very reasonable sounding assumptions -- that quantum effects are confined to within planck distances, and that the vacuum is unique -- then we can establish that there {\it cannot} be any such change to the evolution of fig.\ref{matffourt}.

String theory resolves the problem by telling us that the first assumption is false: quantum gravity effects are {\it not} confined to a given distance, but instead {range over distances that increase with the number of quanta making up the bound state corresponding to the hole}. We find an effect called `fractionation' which shows that in a bound state of strings and branes the quantum effects stretch to distances of order horizon scale (eq.\ref{matqqq}). This is a crude estimate, 
but we can then return to simple black hole states and construct them explicitly, finding in each case that there is no horizon; instead the interior of the hole is a `fuzzball'.

The information paradox was important because its resolution would 
have to challenge some basic assumptions that we have held about quantum gravity. We do indeed  find a change in our basic idea of how quantum gravity acts when we have large dense systems of strings and branes. The goal is now to formalize this understanding and apply it to other basic problems like the early Universe where quantum gravity can be important.

\section*{Acknowledgements}
I would like to thank the organizers for a wonderful school at Mytiline. I am grateful to Steve Avery, Borun Chowdhury and Jeremy Michelson for many helpful comments on this manuscript. This work was supported in part by DOE grant DE-FG02-91ER-40690.


\begin{thebibliography}{99}


\bibitem{matbek}
  J.~D.~Bekenstein,
  Phys.\ Rev.\  D {\bf 7}, 2333 (1973).

\bibitem{mathawking}
  S.~W.~Hawking,
  Commun.\ Math.\ Phys.\  {\bf 43}, 199 (1975)
  [Erratum-ibid.\  {\bf 46}, 206 (1976)];
  S.~W.~Hawking,
  Phys.\ Rev.\  D {\bf 14}, 2460 (1976).
  
   \bibitem{matfuzz}
   S.~D.~Mathur,
  Fortsch.\ Phys.\  {\bf 53}, 793 (2005)
  [arXiv:hep-th/0502050];
  S.~D.~Mathur,
  Class.\ Quant.\ Grav.\  {\bf 23}, R115 (2006)
  [arXiv:hep-th/0510180];
  I.~Bena and N.~P.~Warner,
  arXiv:hep-th/0701216.
  
  \bibitem{matbirrel}
  N.~D.~Birrell and P.~C.~W.~Davies,
  ``Quantum Fields In Curved Space,''
{\it  Cambridge, Uk: Univ. Pr. ( 1982) 340p;}
  S.~A.~Fulling,
  ``Aspects of quantum field theory in curved space-time,"  London Math.\ Soc.\ Student Texts {\bf 17}, 1 (1989).

  
  \bibitem{matwald}
  R.~M.~Wald,
  Commun.\ Math.\ Phys.\  {\bf 45}, 9 (1975).
  
  \bibitem{matparker}
  L.~Parker,
  Phys.\ Rev.\  D {\bf 12}, 1519 (1975).

\bibitem{matgiddings}
  S.~B.~Giddings and W.~M.~Nelson,
  Phys.\ Rev.\  D {\bf 46}, 2486 (1992)
  [arXiv:hep-th/9204072].
  
   
  \bibitem{matstretch}
  S.~D.~Mathur,
  Int.\ J.\ Mod.\ Phys.\  A {\bf 15}, 4877 (2000)
  [arXiv:gr-qc/0007011];
  S.~D.~Mathur,
  Int.\ J.\ Mod.\ Phys.\  D {\bf 11}, 1537 (2002)
  [arXiv:hep-th/0205192].
  
   \bibitem{matmc}
  A.~Chamblin and J.~Michelson,
  Class.\ Quant.\ Grav.\  {\bf 24}, 1569 (2007)
  [arXiv:hep-th/0610133].

  
   \bibitem{matsen}
L.~Susskind,
arXiv:hep-th/9309145;
J.~G.~Russo and L.~Susskind,
Nucl.\ Phys.\ B {\bf 437}, 611 (1995)
[arXiv:hep-th/9405117];
%
A.~Sen,
Nucl.\ Phys.\ B {\bf 440}, 421 (1995)
[arXiv:hep-th/9411187];
A.~Sen,
Mod.\ Phys.\ Lett.\ A {\bf 10}, 2081 (1995)
[arXiv:hep-th/9504147];
%
C.~Vafa,
  Nucl.\ Phys.\  B {\bf 463}, 435 (1996)
  [arXiv:hep-th/9512078].


\bibitem{matwi}
  R.~M.~Wald,
  Phys.\ Rev.\  D {\bf 48}, 3427 (1993)
  [arXiv:gr-qc/9307038].


   \bibitem{matdabholkar}
  A.~Dabholkar,
  Phys.\ Rev.\ Lett.\  {\bf 94}, 241301 (2005)
  [arXiv:hep-th/0409148].
  
   \bibitem{matsv}
A.~Strominger and C.~Vafa,
Phys.\ Lett.\ B {\bf 379}, 99 (1996)
[arXiv:hep-th/9601029];
%
C.~G.~.~Callan and J.~M.~Maldacena,
Nucl.\ Phys.\ B {\bf 472}, 591 (1996)
[arXiv:hep-th/9602043].
%

 \bibitem{matstring}
   O.~Lunin and S.~D.~Mathur,
  Nucl.\ Phys.\  B {\bf 623}, 342 (2002)
  [arXiv:hep-th/0109154];
   O.~Lunin and S.~D.~Mathur,
  Phys.\ Rev.\ Lett.\  {\bf 88}, 211303 (2002)
  [arXiv:hep-th/0202072].



  \bibitem{matmany}
  V.~Balasubramanian, J.~de Boer, E.~Keski-Vakkuri and S.~F.~Ross,
Phys.\ Rev.\ D {\bf 64}, 064011 (2001), hep-th/0011217;
J.~M.~Maldacena and L.~Maoz,
JHEP {\bf 0212}, 055 (2002)
[arXiv:hep-th/0012025];
 S.~Giusto, S.~D.~Mathur and A.~Saxena,
  Nucl.\ Phys.\  B {\bf 710}, 425 (2005)
  [arXiv:hep-th/0406103];
  O.~Lunin,
JHEP {\bf 0404}, 054 (2004)
[arXiv:hep-th/0404006].
I.~Bena and N.~P.~Warner,
  arXiv:hep-th/0701216;
 I.~Bena and N.~P.~Warner,
  Adv.\ Theor.\ Math.\ Phys.\  {\bf 9}, 667 (2005)
  [arXiv:hep-th/0408106];
  V.~Balasubramanian, E.~G.~Gimon and T.~S.~Levi,
  arXiv:hep-th/0606118;
 I.~Kanitscheider, K.~Skenderis and M.~Taylor,
  arXiv:0704.0690 [hep-th].

\bibitem{matross}
  V.~Jejjala, O.~Madden, S.~F.~Ross and G.~Titchener,
  Phys.\ Rev.\  D {\bf 71}, 124030 (2005)
  [arXiv:hep-th/0504181].
  
\bibitem{matmyers}
  V.~Cardoso, O.~J.~C.~Dias, J.~L.~Hovdebo and R.~C.~Myers,
  Phys.\ Rev.\  D {\bf 73}, 064031 (2006)
  [arXiv:hep-th/0512277].

  \bibitem{matcm}
  B.~D.~Chowdhury and S.~D.~Mathur,
  arXiv:0711.4817 [hep-th].

  
  \bibitem{matdmpre}
  S.~R.~Das and S.~D.~Mathur,
  Phys.\ Lett.\  B {\bf 375}, 103 (1996)
  [arXiv:hep-th/9601152].
  
   
  \bibitem{matmaldasuss}
J.~M.~Maldacena and L.~Susskind,
Nucl.\ Phys.\ B {\bf 475}, 679 (1996)
[arXiv:hep-th/9604042].

  
\bibitem{matemission}
S.~D.~Mathur,
Nucl.\ Phys.\ B {\bf 529}, 295 (1998)
[arXiv:hep-th/9706151].
%

  
  \bibitem{matpreskill}
  T.~M.~Fiola, J.~Preskill, A.~Strominger and S.~P.~Trivedi,
  Phys.\ Rev.\  D {\bf 50}, 3987 (1994)
  [arXiv:hep-th/9403137];
  E.~Keski-Vakkuri and S.~D.~Mathur,
  Phys.\ Rev.\  D {\bf 50}, 917 (1994)
  [arXiv:hep-th/9312194].
  
 
   \bibitem{matweb}
  S.~D.~Mathur,
  arXiv:0706.3884 [hep-th].
  



\end{thebibliography}
\end{document}